\documentclass[article,amsmath,11pt,amssymb,floatfix,superscriptaddress,
showpagenumber,nofootinbib,showpacs,showkeys,prd,onecolumn]{revtex4}
\usepackage{graphicx}
\usepackage{epsfig}
\usepackage{bm}
\usepackage{amsfonts}
\usepackage{xcolor}
\usepackage{amsmath}
\usepackage{amssymb}
\usepackage{lineno}
\usepackage{xspace} 
\usepackage{threeparttable}
\usepackage{dcolumn}
\input epsf

\begin{document}
\title{\Large Compact objects in AdS spacetime with exponential, quadratic and power-law bosonic mass profiles}

\author{Samprity Das}
\email{samprity.das@s.amity.edu}
\affiliation{ Department of Mathematics, Shibpur Dinobundhoo Institution (College),\\
Shibpur, Howrah 711102; \\
Department of Mathematics, Amity University, Kolkata, New Town, Rajarhat, Kolkata 700135,
India.}

\author{Aroonkumar Beesham }
\email[Email: ]{abeesham@yahoo.com}
\affiliation{  Department of Mathematical Sciences, University of Zululand, P Bag X1001, Kwa-Dlangezwa 3886, South Africa, \\
 Faculty of Applied and Health Sciences, Mangosuthu University of Technology, P O Box 12363, Jacobs 4026, South Africa, \\
National Institute for Theoretical and Computational Sciences (NITheCS), South Africa, \\
DSTI-NRF Centre of Excellence in Mathematical and Statistical Sciences (CoE-MaSS), South Africa.}

\author{Surajit Chattopadhyay}
\email{schattopadhyay1@kol.amity.edu; surajitchatto@outlook.com}
\affiliation{ Department of Mathematics, Amity University, Major
Arterial Road, Action Area II, Rajarhat, New Town, Kolkata 700135,
India.(Communicating author)}

\date{\today}

\begin{abstract}
This paper reports a study on the formation and physical characteristics of compacts stars in AdS spacetime within the framework of Bose-Einstein Condensate. Considering a Bose-Einstein condensate background at zero temperature this study works on total mass, compactness, surface redshift, density, pressure, adiabatic index and energy conditions. The bosonic mass has been taken as three distinct functions of radial coordinate in exponential form, quadratic form, and power law form. Our results reveal that the mass increases monotonically with radius and remains within observational limit for all the observationally motivated compact-star mass scales considered in this study and the compactness for all the cases is within Buchdahl's limit and hence it was confirmed that the configuration correspondence to compact stellar configuration models rather than forming a collapsing model. Both NEC and SEC are satisfied throughout the stellar interior and hence dynamical stability is ensured. Furthermore, the study also confirms the enhanced mass concentration near the outer region in the stellar models under consideration. Hence present study explores the physical properties and stability of compact bosonic configurations in AdS spacetime within a holographically motivated framework.

The present analysis is primarily phenomenological and qualitative in nature. The models considered here are intended to explore possible behaviours of self-gravitating bosonic configurations in AdS geometry and are not proposed as fully realistic neutron-star models.

\pacs {04.50.Kd; 97.60.Jd}
\keywords {Anti De-sitter Space, Compact Objects, Bosonic mass, Boson energy, Surface Redshift, Stability Analysis. }
\end{abstract}

\maketitle
\section{Introduction}

Compact objects are natural laboratories to study the exotic matter present in the universe. White dwarfs, black holes, and compact stellar configurations are examples of astrophysical compact objects that are created by gravitational collapse. These classes take into account their inner density and mass \cite{shapiro2024black}. Some of these tiny objects, such Her X-1 (X-ray pulsar), PSR 0943+10, 4U 1820-30 (X-ray burster), 4U 1728-34, RX J185635-3754 (X-ray sources), and SAX J 1808.4-3658 (millisecond pulsar), clearly imply that these small objects may be strange stars. The presence of (hyper)nuclear matter in compact stars may exhibit several remarkable properties, such as superfluidity and superconductivity, a contained neutrino component in the early stages of evolution, or extraordinarily high magnetic fields. In March 2010, a team conducted a dense sequence of observations of J1614–2230 using the National Radio Astronomy Observatory Green Bank Telescope (GBT) \cite{sedrakian2023heavy, shapiro1983physics, weber2017pulsars}. The idea that everything in the universe occurs on a 4-dimensional Riemannian geometry known as spacetime and that matter and curvature are related through the use of Einstein's field equations (EFE) forms the basis of Einstein's theory of gravity \cite{einstein1916foundation, lorentz1952principle, dirac1996general}, which has expanded our understanding of astronomical objects and accelerated the universe's expansion. Using the energy-momentum tensor $T_{\mu\nu}$, different types of fields will characterize the matter composition of the space-time. 

In 1998, Maldacena found the duality between gravity and gauge theory in spacetime with different dimensionalities \cite{maldacena1999large}.  It is hypothesized that a gauge theory in four-dimensional Minkowski spacetime $(M_4)$ is dual to the $type-IIB$ string theory in $AdS_5 \times S_5$ at the frontier of the AdS space. By avoiding the uncontrollable non-perturbative calculation by the use of weak-strong duality, the correspondence can be utilized as a supplemental approach to explore the strongly coupled gauge theory in four-dimensional Minkowski spacetime, a cousin of quantum chromodynamics \cite{burikham2014comments}. This problem can be detoured by the calculation in the tractable weakly interacting string theory in five dimensional Anti de
Sitter space $(AdS_5)$. String theory comprises all kinds of fundamental particles and forces, and thus in principle can be used to describe all forms of matter. A recent hot topic in theoretical astrophysics is holographic duality. ’t Hooft \cite{hooft1993dimensional} suggested that the lowering of dimension at the horizon reveals the quantum gravity effect.  According to this theory, a region of space can only hold a certain amount of energy before gravitational collapse takes place. Because the area of a black hole's event horizon determines its entropy, gravitational collapse places a limit on the amount of information that can be stored there. This maximum entropy is given by that of a black hole occupying the entire region and is proportional to the horizon area measured in units of the Planck area. In this sense, all the information contained within a region of space can be encoded on its boundary, leading to a holographic description. Author in \cite{witten1998anti} has elaborately explained the correspondence between AdS space and holography.

As compact stellar configurations involve highly dense effective matter distributions, they are not asymptotically dense, and the robust coupling characteristic of the problem makes the AdS/CFT correspondence a viable theoretical tool, it is possible to study compact stellar configurations in AdS space and holographic correspondence.  \cite{kovensky2022building, kovensky2022predictions}
Although a rigorous strong-coupling computation is made possible by the holographic principle, \cite{maldacena1999large,witten1998anti} analysis is nowadays and in the near future limited to string models whose counterpart is essentially different from the pertinent underlying field theory, quantum chromodynamics (QCD). Because they have the potential to collapse in their gravitational field and generate a black hole, compact stellar configurations and other types of compact stellar configurations, such as quark stars and Strage stars, are the predecessors of black holes. Three thermal phases, including pure radiation, small black hole (SBH), and large black hole (LBH), are known to exist in the AdS. The deconfinement phase transition of the confined gauge matter to the quark-gluon plasma (QGP) on the boundary is dual to the Hawking-Page transition of the pure radiation to the black hole in the bulk AdS space.Holographically, the gas of confined hadrons is represented by the radiation in the thermal AdS, whereas the gas of deconfined quarks and gluons, which we shall simply refer to as the QGP from now on, is represented by the radiation in the AdS-BH background. Gravitational collapse-induced classical transition is a widespread phenomenon that is not exclusive to pure radiation. We can think of AdS stars with any number of particles of bosons \cite{de2010holographic}, fermions to study the collapse \cite{arsiwalla2011degenerate}.
Multi-trace composite operators in the CFT side as the dual of the bulk fermionic star were successfully constructed by the authors of a groundbreaking study on the holographic degenerate star with fermion content in the work \cite{de2010holographic}. Holographically, the thermalization of the gauge matter on the boundary is dual to the gravitational collapse of an AdS star \cite{shuryak2005gravity,chesler2008horizon,bhattacharyya2009weak,balasubramanian2011thermalization}. 

The most important characteristic of bosonic systems is their phase transition to a condensed state. Because every particle in this kind of quantum bosonic system is in the same quantum ground state, it is referred to as a Bose-Einstein Condensate (BEC). Such a system in both coordinate and momentum space is physically characterized by a prominent peak across a larger spread. The Bose-Einstein condensation process is thought to be largely responsible for the understanding of many basic condensed matter physics processes.  Since Bose-Einstein condensation has been observed and thoroughly studied in Earthly systems, the possibility that it may also exist in bosonic systems happening at the astrophysical or cosmic scales cannot be instantly ruled out \cite{das2025relativistic}. Therefore, the probability of Bose Einstein condensate existing as Dark Matter exists \cite{ji1994late,hu2000fuzzy}. The possible presence of Bose Einstein condensate in astrophysical objects has been a foundation for recent research \cite{das2025relativistic,li2010j,robles2012flat,chavanis2011mass}.

In simple terms AdS space has timelike boundary at infinity, which acts like a physical wall or in alternative words we can suggest it acts a gravitational box. This confined box prevents information or pulls the matter back rather than escaping to infinity \cite{witten1998anti}. If we think of the infinite boundary as a screen that encodes all the information supporting the holographic principle, then this condition naturally supports the holographic principle as the light ray or the fields reflect back off the boundary. The relationship between two-dimensional conformal field theory and three-dimensional Chern-Simons gauge theory is also evocative of the realization of holography via AdS space \cite{witten1989quantum}. Based on the well-known AdS/CFT duality, which is a strong/weak duality in that the gravitational theory is weakly curved when the effective coupling of the gauge is large, holography is an impressive non perturbative technique that can be used to describe some strongly interacting matter, like compact stellar configurations \cite{feng2024holographic}. The AdS/CFT correspondence, or holography, is a useful theoretical tool because of the strong coupling nature, which is that compact stellar configuration matter is dense but not asymptotically dense \cite{maldacena1999large,kovensky2022building}. This area of compact stellar configuration research has significant implications for astrophysics. Because of their extremely high core densities, these astrophysical objects are regarded as natural laboratories for researching ultrahigh density matters that are not achievable in the earthly system. Neutron star observations have historically depended on detecting electromagnetic (EM) radiation from pulsars or measuring the characteristics of binary systems, in which one component is a compact stellar configuration and the other, for example, a white dwarf. Numerous characteristics of compact stellar configurations, including as their masses, radii, surface temperatures, and spinning rates, can be deduced from these data. Numerous studies have been conducted to support the study of compact stellar configurations in AdS/CFT duality or holography while examining representative compact-object configurations \cite{kovensky2022building, hoyos2022holographic,feng2024holographic}. 

The present work does not attempt to model astrophysical reference compact-object configurations directly. Rather, compact bosonic configurations in AdS spacetime are studied as theoretical gravitational systems motivated by holographic considerations. The comparison with observed compact-star mass scales is used only as a qualitative reference scale and not as evidence that astrophysical compact stellar configurations exist in AdS spacetime.

Before moving to the objective of our work it is important to compare our findings with the previous study on compact object in the AdS configuration such as \cite{de2010holographic,burikham2014comments, arsiwalla2011degenerate}. In this studies we get to see that the authors explore the compact objects for degenerate Fermi gas background in AdS space. Bulk parameters, including sound speed, adiabatic index, and entropy density within the star, are computed both analytically and numerically. Authors have also confirmed that in the hydrodynamic limit, the Fermi gas constitutes a degenerate star with a radius dictated by the Fermi level. In the contrary we have evaluated the compact objects for Boson gas and considered the energy of boson in AdS space as the AdS star functions as a radiation star, which is feasible for a fermionic star or a Bose- Einstein condensate star at zero temperature. In this way our study explores the exotic compact objects with self interacting particles at the core. Also, our study consists of three distinct forms of bosonic mass, which we are going to consider as a function of radial coordinate and explored the physical attributes of the compact objects such as density, pressure and mass.

The objective of the present analysis is to investigate the qualitative behavior and stability of self-gravitating bosonic matter distributions in AdS geometry, rather than to construct fully realistic compact-object models. Using the bulk density of state for boson total mass of the compact object has been evaluated in terms of bosonic energy and bosonic mass in $(d-2)$ dimension. In Section II, we have described brief overview in AdS space, and evaluated number of particles and total mass within a compact object. In Section III, we have considered the bosonic mass as function of radial coordinate in three distinct forms: as a power law function of radius, an exponential function, and a quadratic function. Three distinct scenarios have been thoroughly examined, each involving the physical characteristics of the star. The stability analysis, energy conditions have also been verified for compact astrophysical configurations such as \emph{LMC X-4, PSR J0740+6620, PSR J1614-2230, PSR J0348+0432, PSR J0030+0451, and EXO 1785-248} for the above mentioned subcases. It may be noted that the observational compact-object data are used only as representative astrophysical scales for comparison of mass and compactness, and not as direct evidence for AdS bosonic stars. In the present work, the AdS bosonic configurations are considered as theoretical self-gravitating systems motivated by holographic ideas, and not as realistic neutron star models. The observational compact-object data used in this paper are taken only as reference values for mass and compactness for qualitative comparison. The study does not intend to claim that the observed neutron stars exist in AdS spacetime or that they are actual bosonic condensate stars. The main aim of this work is to study how different bosonic mass distributions affect the compactness, stability, and thermodynamic behaviour within an AdS-inspired gravitational framework. In the present study, the chosen bosonic mass profiles are taken as phenomenological assumptions to study different possible behaviours of compact bosonic configurations in AdS spacetime. These profiles are not obtained from any fundamental microscopic field theory or from a complete Einstein–Klein–Gordon model. The main purpose is to examine how different radial mass distributions affect the physical properties and stability of the configurations within the considered framework. The final concluded remark have been summarized in Section IV.

\section{Compact star overview in AdS space}

 The Conformal field theory (CFT) is defined on the cylinder $R \times S^{d-1}$ with the metric
\begin{equation}
ds^2= -dt^2+l^2 d\Omega^2 _{d-1},
    \label{E40}
\end{equation}
for $d\Omega^2_{d-1}$ is the standard mertic on $S^{d-1}$. The cylinder is the boundary of the Anti de-Sitter $AdS_{d+1}$ with $l$ as the curvature radius. Now, we conformally map the cylinder on the euclidean plane $R^d$ by continuing to analytically compute euclidean time, $ t \rightarrow i l log |x|.$
The angular coordinate is defined by $\Omega$. The states of CFT correlate to operators through
\begin{equation}
\lim_{|x|\to 0} \Phi(x)\,|{\rm vac}\rangle = |\Phi\rangle
    \label{E41}.
\end{equation}
We now assume that the CFT has a large $c$ limit where single and multi trace operators can be distinguished, and that it contains both bosonic and fermionic fields. In our study we are going to consider the bosonic fields. The Einstein equations can be achieved by varying the action with metric and the matter field \cite{hartmann2013compact}, 
\begin{equation}
G_{\mu\nu}=R_{\mu\nu}-\frac{1}{2}g_{\mu\nu} (R-2\Lambda)=8\pi G T_{\mu\nu},
    \label{E42}
\end{equation}
with stress-energy tensor 
\begin{equation}
T_{\mu\nu}=g_{\mu\nu}L_M-2\frac{\partial L_M}{\partial g^{\mu\nu}},
    \label{E43}
\end{equation}
and the matter field equation 
\begin{equation}
\Delta_\mu \Delta^\mu \Phi=-\lambda \frac{\Phi}{|\Phi|},
    \label{E44}
\end{equation}
where $\Delta _\mu$ is covariant derivative. The global phase transition is invariant under the action 
\begin{equation}
    \Phi \rightarrow \Phi e^{i x},
\end{equation}
which leads to the current conservation as 
\begin{equation}
    j^\mu=-i (\Phi^* \partial^\mu \Phi- \Phi \partial^\mu \Phi ^*).
\end{equation}
As a result, the bosonic field property is extended in the next sections, and the AdS/CFT theory with a bosonic field background is studied.

\subsection*{Hydrodynamic Description}

To evaluate spherically symmetric star we apply Schewrzchild like coordinates and let us write the metric for AdS in the form
\begin{equation}
    ds^2=g_{\mu\nu}dx^\mu dx^\nu=-A^2(r) dt^2+(A^2(r))^{-1} dr^2+r^2d\Omega ^2_{d-2},
    \label{E01}
\end{equation}
where $A(r)=\sqrt{(1+\frac{r^2}{l^2})}$ and $l$ is AdS radius. For standard hydrodynamic expression the stress energy tensor of boson in the form
\begin{equation}
    T_{\mu\nu}=(\rho+p)u_{\mu}u_{\nu}+pg_{\mu\nu},
    \label{E02}
\end{equation}
where $u_{\mu}$ is a static velocity field: $u_{\mu}dx^\mu=A(r)dt$. The radial density $\rho$ and pressure $p$ are determined by stress energy conservation, which is \cite{de2010holographic}
\begin{equation}
    \frac{dp}{dr}+\frac{1}{A}\frac{dA}{dr}(\rho+p)=0,
    \label{E03}
\end{equation}
and the coupled equation between mass ($M$) and chemical potential ($\mu$) obtained as
\begin{equation}
M'(r)=V_{d-2}\rho (r) r^{d-2},
    \label{E04}
\end{equation}
\begin{equation}
\mu '(r)=\mu (r)\left(\frac{B'(r)}{B(r)}-\frac{V_{d-2}C_{d-1}}{2}\left(\rho (r)c^2+p(r)\right)+r B^2(r)\right),
    \label{E05}
\end{equation}
where $B(r)=(A^2(r))^{-1}$, $V_{d-2}=$ area of a unit (d-2) sphere, $C_{d-1}=\frac{16 \pi G}{(d-2) V_{d-2}c^4}$=constant. For AdS space, chemical potential with respect to radial coordinate
\begin{equation}
\mu = \frac{\epsilon_B}{A(r)},
    \label{E06}
\end{equation}
where $\epsilon_B$ is the energy of boson and central chemical potential for empty AdS with boson is $A(0)=1$.

For a boson star, at null temperature, the density and pressure of the boson of the star are given
\begin{equation}
    \rho=\frac{g_B V_{(d-2)}}{(2 \pi)^{(d-1)}} \int _m^{\mu } \mu^2 (\mu^2-m_B ^2)^{(d-3)/2} d\mu  ,
\end{equation}
\begin{equation}
    p=\frac{g_B V_{(d-2)}}{(d-1)(2 \pi)^{(d-1)}} \int _m^{\mu } (\mu^2-m_B ^2)^{(d-1)/2} d\mu .
\end{equation}
Here $g_B$ is the number of internal degrees of freedom i.e. spin, possible color of boson etc. For, $m_B=0$, is noted as conformal limit, the density and pressure become $\rho=\frac{g_B V_{(d-2)}}{(2 \pi)^{(d-1)}} \frac{\mu ^d}{d}$ , $p=\frac{\rho}{d-1}$, which is notes as linear equation of state \cite{burikham2014comments}. In case of conformal limit i.e. for $m=0$, the AdS star behaves like a radiation star, which explains at zero temperature feasiable scenario is for a fermionic star or a Bose-condensate star. The temperature distribution of the radiation star follows the relation $T(r)=T(0)/A(r)$. For the purpose of analyzing the AdS star, we assume zero temperature. Temperature can only introduced when comparing the AdS black hole (AdS-BH) with thermal AdS. 

Now, Boson number density($n$) is equal to the number of particles per unit volume for ($d-2$)-dimension
\begin{equation}
n=\frac{V_{d-2}}{(d-1)(2\pi )^{d-1}}\left(P_B\right){}^{d-1} ,
    \label{E07}
\end{equation}
where $P_B$= boson momentum. Now, boson momentum related to chemical potential ($\mu$) as
\begin{equation}
\mu =\sqrt{P_B{}^2+m_B{}^2} ,
    \label{E08}
\end{equation}
where $m_B$= mass of boson gas, which is described in relativistic case. Also, boson energy establish an relation with boson momentum in non-relativistic case such as
\begin{equation}
\epsilon_B=\frac{P_B^2}{2 m_B} .
    \label{E09}
\end{equation}
Now to avoid this contradictory scenario we chose the energy equation of free boson particles in relativistic model \cite{wachter2011relativistic}
\begin{equation}
E^2= m_B^2c^4+P_B^2c^2.
    \label{Eq1}
\end{equation}
As, $P_B <<m_B$ and $c^2=1$ give $E= m_B \sqrt{1+\frac{P^2}{m^2}}$, which leads to $\epsilon_B=(E-m)= \frac{P_B^2}{2m}$ as stated in Eq. (\ref{E09}).
The bulk density of state for boson with spatial ($d-2$) dimension is \cite{bhuiyan2021bose}
\begin{equation}
g(\epsilon_B )=V_{d-2}\left(\frac{m_B}{2 \pi  \hbar }\right)^{\frac{d-2}{2}}\frac{(\epsilon_B -a n)^{\frac{d-4}{2}}}{\Gamma \left(\frac{d-2}{2}\right)} ,
    \label{E10}
\end{equation}
where $\hbar$= Planck's constant and '$a$' is positive constant. 
Hence, using the bulk density of the boson in Eq. (\ref{E10}), we can evaluate the number of particles and the total mass within a compact star such as 
\begin{equation}
\begin{array}{cc}
   N=\int _m^{\epsilon_B }g(\epsilon_B )d\epsilon  \\
    =\left(\frac{m_B}{2 \pi  \hbar }\right)^{\frac{d-2}{2}}\frac{V_{d-2}}{\Gamma \left(\frac{d-2}{2}\right)}\left(\frac{2(\epsilon_B -{an})^{\frac{d-2}{2}}}{d-2}-\frac{2(m_B-{an})^{\frac{d-2}{2}}}{d-2}\right) ,
\end{array}
    \label{E11}
\end{equation}
and
\begin{equation}
\begin{array}{cc}
   M=\int _m^{\epsilon_B }\epsilon_B  g(\epsilon_B )d\epsilon \\
    = \left(\frac{m_B}{2 \pi  \hbar }\right)^{\frac{d-2}{2}}\frac{V_{d-2}}{\Gamma \left(\frac{d-2}{2}\right)}\frac{2}{d(d-2)}\left(2 a n\left((\epsilon_B
-{an})^{\frac{d-2}{2}}-(m_B-{an})^{\frac{d-2}{2}}\right) \right)+ \\
 \left(\frac{m_B}{2 \pi  \hbar }\right)^{\frac{d-2}{2}}\frac{V_{d-2}}{\Gamma \left(\frac{d-2}{2}\right)}\frac{2}{d(d-2)} \left( (d-2) \left((\epsilon_B (\epsilon_B -{a n})^{\frac{d-2}{2}}-m_B(m_B-{an})^{\frac{d-2}{2}}\right)\right) ,
\end{array}
    \label{E12}
\end{equation}
respectively. For simplicity of our study we have considered $\hbar=1$ in all of the aforementioned equations hereafter. 

\section{Mass of the Compact stars}

By using the properties of gamma function 
\[
x\Gamma(x) = \Gamma(x+1),
\qquad
\Gamma(x)\Gamma\left(x+\frac{1}{2}\right)
= 2^{1-2x}\sqrt{\pi}\,\Gamma(2x)
\]
and $V_{d-2}=2 \pi ^{(d-1)/2}/ \Gamma((d-1)/2)$ the mass of the compact stars can be calculated from Eq. (\ref{E12}). We consider $d=5$ and get the mass of compact star in AdS$_5$ in the form of from Eq. (\ref{E12})
\begin{equation}
\begin{array}{cc}
   M=m_B{}^{3/2}\mu \left(\frac{48 \pi ^2\frac{\epsilon _B}{\mu }+a \mu ^3\left(\frac{m_B{}^2}{\mu ^2}-1\right)^2}{60 \sqrt{2}\pi
^2}\right)\left(\mu ^{3/2}\left(\frac{\epsilon _B}{\mu }-\frac{{a\mu}^3\left(\frac{m_B{}^2}{\mu ^2}-1\right)}{32 \pi ^2}\right)^{3/2}-\left(m_B-\frac{{a\mu}^4\left(m_B{}^2-\mu ^2\right)}{32 \pi ^2}\right)^{3/2}\right) .
\end{array}
    \label{E13}
\end{equation}
Now, chemical potential $\mu$ is a very small quantity and also for higher temperature $\mu \rightarrow 0$. Hence higher orders terms of $\mu$ in Eq. (\ref{E13}) can be neglected and which gives more simplified form of the mass for the compact stars as
\begin{equation}
   M=m_B{}^{3/2}\mu  \times \frac{2\sqrt{2}}{5} \sqrt{1+\frac{r^2}{l^2}}\left(\mu ^{3/2}\left(1+\frac{r^2}{l^2}\right)^{3/2}-m_B^{3/2}\right) .
    \label{E14}
\end{equation}
Relation between bosonic mass and chemical potential can be established using Eq. (\ref{E08}) such as $\mu$ and $m_B$ are in direct relation represented as 
\begin{equation}
    \mu= \alpha m_B ,
    \label{E15}
\end{equation}
hence $(\alpha^2-1)m_B^2=P_B^2$. For positive momentum, $P_B^2 >0$ $\implies$ $|\alpha| >1$. 

Thus the mass of the compact stars reformulated from Eq. (\ref{E14}) as
\begin{equation}
M= \frac{2}{5}\sqrt{2}m_B{}^4\alpha  \sqrt{1+\frac{r^2}{l^2}}\left(\alpha ^{3/2}\left(1+\frac{r^2}{l^2}\right)^{3/4}-1\right)
    \label{E16}
\end{equation}

For a finite constant $l$, the AdS/CFT correspondence holds. We will now set $l=1$ and examine whether our research results in quark-gluon plasma with a negative heat capacity. For a certain $m_B$, increasing $l$ will raise the bosonic star's mass limit; but, we can always increase $m_B$ to make the mass limit much smaller, even smaller than $m_B$.

Now, we consider the bosonic mass ($m_B$) as function of radial coordinate $r$ in different forms such as:

$(i)$ $m_B$ as exponential function of radius,

$(ii)$ $m_B$ as quadratic function of radius,

$(iii)$ $m_B$ as power law form of radius.

In a number of astrophysical modeling applications, the quadratic, linear, and exponential forms are well-debated premises. For example, the authors have selected a quadratic form for the radial pressure under modified gravity \cite{bhar2017compact}. In the paper, the authors used a shape function in the exponential form of radius \cite{moraes2019exponential}. In their works, authors have also used the density in quadratic, linear, and exponential forms \cite{das2023modified,das2023study}. We used the quadratic form of radial coordinates in our modeling, drawing inspiration from these studies.

\subsection*{Mass function as an exponential form of radius}

Considering bosonic mass as an exponential function of the radial coordinate 
\begin{equation}
m_B=\gamma e^{\eta r}+ \delta e^{- \eta r} ,
    \label{E17}
\end{equation}
where $\eta$ is an arbitrary constant. Consequently, the modified mass for compact stars is expressed as
\begin{equation}
M= \frac{2}{5} \sqrt{2}\, e^{-4r\eta} \sqrt{1+r^{2}} \left( e^{2r \eta} \gamma + \delta \right)^{4} \alpha  \left( -1 + (1+r^{2})^{3/4} \alpha^{3/2} \right)
 \label{E18}
\end{equation}
Moreover, the compactness of a pulsar and, thus, it's physical nature can be inferred from the mass-radius ratio \(u = m/r\). The greatest value of the mass-radius ratio for isotropic matter distributions is limited by \(u < 4/9\) in accordance with Buchdahl's limit \cite{das2025compact}. Hence, modified compactness for our model is
\begin{equation}
 u= \frac{2 \sqrt{2} \, e^{-4r\eta} \sqrt{1 + r^2} \left(e^{2r\eta} \gamma + \delta \right)^4 \alpha \left(-1 + (1 + r^2)^{\tfrac{3}{4}} \alpha^{\tfrac{3}{2}}\right)}{5r} .
\label{E32}
\end{equation}
The mass-radius ratio hypothesis usually classifies compact objects in the following ways: \\
$(i)$ Regular Star :($u \sim 10^{-5}$),
$(ii)$ White Dwarf : ($u \sim 10^{-3}$),
$(iii)$ Neutron Star : ($0.1<u<0.25$),
$(iv)$ Ultra Compact Object : ($0.25<u<0.5$),
$(v)$ Black Hole : ($u=0.5$) . 
We have developed some representative compact-object models perturbing the parameter $\alpha$ as represented in Table \ref{T1}.

\begin{table}[ht]
\centering
\begin{threeparttable}
\caption{From Eq. (\ref{E18}) the model mass, compactness, and surface redshift for specific pulsars have been evaluated for the exponential model}
\label{T1}
\begin{tabular}{clccccc}
\hline
\textbf{Sl. No.} & \textbf{Star Model} & 
\textbf{Mass ($M/M_{\odot}$)} & 
$\boldsymbol{\alpha}$ & 
\textbf{Model Mass ($M_{\odot}$)} & 
\textbf{Compactness} & 
\textbf{Surface Redshift} \\
\hline

1 & LMC X-4 & $1.04 \pm 0.09$ & 1.20 & 1.10 & 0.129412 & 0.161554 \\

2 & PSR J0740+6620 & $2.07 \pm 0.11$ & 1.45 & 2.10001 & 0.170733 & 0.232284 \\

3 & PSR J1614-2230 & $1.908 \pm 0.016$ & 1.50 & 1.82712 & 0.140548 & 0.179409 \\

4 & PSR J0348+0432 & $2.01 \pm 0.04$ & 1.38 & 1.49467 & 0.114975 & 0.139568 \\

5 & PSR J0030+0451 & $1.44 \pm 0.16$ & 1.30 & 1.35927 & 0.108742 & 0.130455 \\

6 & EXO 1785-248 & $1.30 \pm 0.20$ & 1.23 & 1.39990 & 0.135922 & 0.175787 \\

\hline
\end{tabular}
\begin{tablenotes}
\footnotesize
\item $M$ denotes the estimated observed mass of the corresponding compact objects.
\end{tablenotes}
\end{threeparttable}
\end{table}

  \subsubsection*{Surface Redshift}
 The surface redshift can be computed using $z=(1-2u)^{-1/2} -1$. The phenomenon of electromagnetic waves or photons leaving a gravitational well (Einstein shift) in physics and GR is known as gravitational redshift, or the Einstein shift in earlier publications \cite{ivanov2002maximum}. The typical result of energy shrinkage in a redshift is a decrease in wave frequency and an increase in wavelength. For compatibility, the surface redshift is within ($z$) $\leq$ 5.211 \cite{das2025evolution}. Hence, evolution of surface redshift for our model
 \begin{equation}
z=-1 + \frac{1}{\sqrt{1 - \frac{4 \sqrt{2} \, e^{-4r\eta} \sqrt{1 + r^2} \left(e^{2r\eta} \gamma + \delta \right)^4 \alpha \left(-1 + (1 + r^2)^{\tfrac{3}{4}} \alpha^{\tfrac{3}{2}}\right)}.{5r}}}.
\label{E33}
 \end{equation}

\begin{figure}[bt]
\centering
\begin{minipage}{0.44\textwidth}
    \centering
    \includegraphics[width=\textwidth]{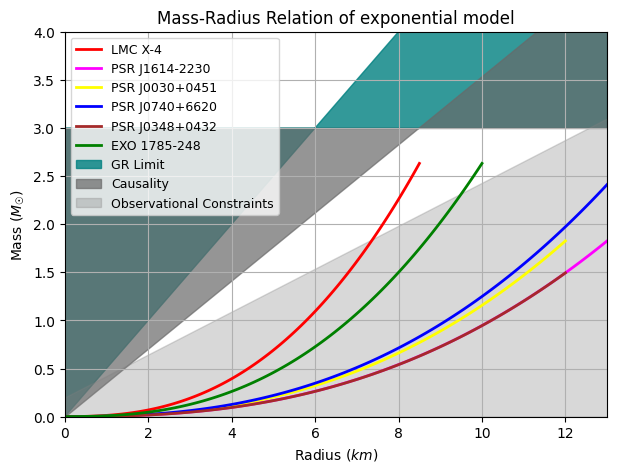}
    \\ \small (a)
\end{minipage}
\hfill
\begin{minipage}{0.44\textwidth}
    \centering
    \includegraphics[width=\textwidth]{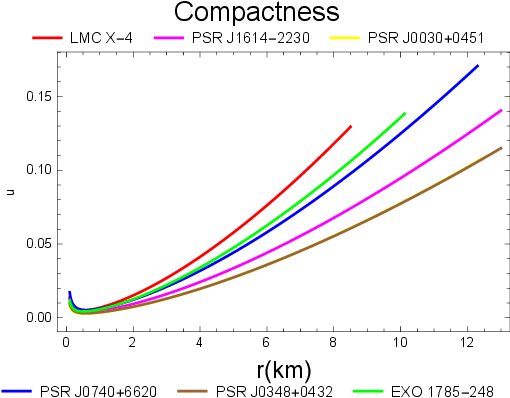}
    \\ \small (b)
\end{minipage}
\hfill
\begin{minipage}{0.44\textwidth}
    \centering
    \includegraphics[width=\textwidth]{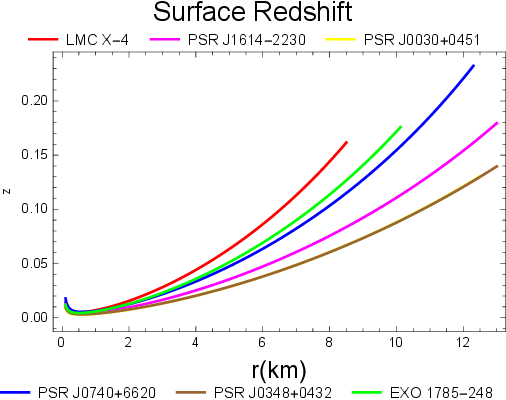}
    \\ \small (c)
\end{minipage}
\caption{(a) Evolution of mass about radius $r$ for exponential model; (b) Evolution of compactness about radius $r$ for exponential model; (c) Evolution of surface redshift about radius $r$ for exponential model.}
\label{f1}
\end{figure}

Fig. \ref{f1} demonstrates the mass, compactness and surface redshift of \emph{LMC X-4, PSR J0740+6620, PSR J1614-2230, PSR J0348+0432, PSR J0030+0451, and EXO 1785-248} using different color grades as mentioned the graphs.
The left panel figure in Fig. \ref{f1} displays the collected mass for each pulsar at the boundary, as well as the mass evolution with respect to the radius, which exhibits a monotonically growing pattern. The compactness about the radius is shown in the figure on the right panel. Here, we can see that compactness of the pulsar decreases and then increases monotonically with decreasing radius. Thus, for larger radius values, the mass-to-radius ratio first decreases and subsequently increases. This suggests that the pulsars have a higher density shell but a lower density core. Furthermore, we may state that the pulsars' mass is expanding sufficiently quickly in relation to their radius close to the core, but it builds up more mass close to the surface. Additionally, the outer layer contains an outer region while the inner core is reasonably stable. However, since the total compactness is smaller than Buchdahl's limit \cite{das2025compact}, we can conclude that pulsars are non-collapsing, meaning that dense objects with limited radius structures can be created without going against general relativity. Additionally, Table \ref{T1} confirms that the chosen pulsars are compact stellar configurations because their compactness falls between 0.1 and 0.25. The lower panel figure depicts the evolution of the surface redshift about radius. We can observe that there is a monotonic increasing pattern in the graphical representation \cite{das2025relativistic,das2025compact,morales2018charged}. Objects have greater gravitational potential, and that implies light loses more energy and time passes more slowly. Hence, it can be concluded that objects are very compact.

\subsubsection*{Physical attributes}

To understand the physical attributes, it is necessary to study the evolution of density and pressure of the compact stellar configuration. Using Eq. (\ref{E04}) we have reformulated the density of the stars for exponential bosonic mass function as follows
\begin{equation}
\begin{split}
\rho &=
\frac{e^{-4 r \eta}\, \alpha \left(e^{2 r \eta} \gamma + \delta\right)^3}
     {5 \sqrt{2}\, \pi^2\, r^3 \sqrt{1 + r^2}} \\
&\quad \times
\Biggl[
\delta \Bigl(
 -8 \eta \bigl(-1 + (1 + r^2)^{3/4} \alpha^{3/2}\bigr)
 - 8 r^2 \eta \bigl(-1 + (1 + r^2)^{3/4} \alpha^{3/2}\bigr)
 + r \bigl(-2 + 5 (1 + r^2)^{3/4} \alpha^{3/2}\bigr)
\Bigr) \\
&\qquad
+ e^{2 r \eta} \gamma \Bigl(
 8 \eta \bigl(-1 + (1 + r^2)^{3/4} \alpha^{3/2}\bigr)
 + 8 r^2 \eta \bigl(-1 + (1 + r^2)^{3/4} \alpha^{3/2}\bigr)
 + r \bigl(-2 + 5 (1 + r^2)^{3/4} \alpha^{3/2}\bigr)
\Bigr)
\Biggr]
\end{split}
\label{E19}
\end{equation}
The number of particles per unit volume within the compact stellar configuration can be evaluated using Eq. (\ref{E07}). For the exponential bosonic mass function, the number of particles per unit volume is reformulated as
\begin{equation}
n=\frac{e^{-4r\eta}(e^{2r\eta } \gamma + \delta)^4(\alpha^2 -1)^2}{34 \pi^4}.
    \label{E20}
\end{equation}
The density and pressure of bosonic stars obey the standard thermodynamic relation in $(d-2)$ dimensions, stated as \cite{arsiwalla2011degenerate}
\begin{equation}
    \begin{array}{ccc}
        d\rho=\mu dn &; & \rho+p=\mu n. 
    \end{array}
    \label{E21}
\end{equation}
Hence, the pressure of the compact stellar configuration can be reformulated using Eq. (\ref{E19}) and Eq. (\ref{E21}) as
    \begin{equation}
\begin{array}{cc}
p=\frac{ e^{-4 r z} (e^{2 r z} x + y)^{3} \alpha}{ 160 \pi^{4} } \Bigg( 
5 e^{-r z} (e^{2 r z} x + y)^{2} ( -1 + \alpha^{2})^{2} \\
- \frac{ 16 \sqrt{2}\, \pi^{2} \Big( 8 (1+r^{2})(e^{2rz}x - y) z 
\big( -1 + (1+r^{2})^{3/4} \alpha^{3/2} \big) 
+ r (e^{2rz}x+y) \big( -2 + 5 (1+r^{2})^{3/4} \alpha^{3/2} \big) \Big)}
{ r^{3} \sqrt{1+r^{2}} }\Bigg) .
    \label{E22} 
\end{array}
\end{equation}
Authors in \cite{das2025relativistic,das2025compact,das2025evolution,das2026anisotropic} have used the observed radius and mass same as the compact objects in this work.

\begin{table}[bt]
\centering
\caption{Evaluation of density, pressure, and number of particles per unit volume for exponential bosonic mass function}
\label{T2}
\begin{threeparttable}
\begin{tabular}{|c|c|c|c|c|c|}
\hline
\textbf{Sl. No.} & \textbf{Star Model} & 
\textbf{Radius (km)} & 
\textbf{Density} & 
\textbf{Pressure} & 
\textbf{Number of Particles/Volume} \\
 & & & $(\mathrm{MeV/fm^3})$ & $(\mathrm{MeV/fm^3})$ & \\
\hline

1 & LMC X-4 & $8.301 \pm 0.2$ 
& $1.3612 \times 10^{17}$ 
& $2.44555 \times 10^{20}$ 
& $3.50978 \times 10^{22}$ \\
\hline

2 & PSR J0740+6620 & $12.34^{+1.89}_{-1.67}$ 
& $4.48861 \times 10^{34}$ 
& $1.59502 \times 10^{37}$ 
& $2.13185 \times 10^{44}$ \\
\hline

3 & PSR J1614-2230 & $13 \pm 2$ 
& $5.76999 \times 10^{34}$ 
& $1.73703 \times 10^{37}$ 
& $2.83493 \times 10^{44}$ \\
\hline

4 & PSR J0348+0432 & $13 \pm 2$ 
& $3.02048 \times 10^{34}$ 
& $1.40825 \times 10^{37}$ 
& $1.36531 \times 10^{44}$ \\
\hline

5 & PSR J0030+0451 & $13.02 \pm 1.24$ 
& $1.75814 \times 10^{34}$ 
& $1.21162 \times 10^{37}$ 
& $7.48639 \times 10^{43}$ \\
\hline

6 & EXO 1785-248 & $10.10 \pm 0.44$ 
& $9.71449 \times 10^{33}$ 
& $1.05387 \times 10^{37}$ 
& $3.91382 \times 10^{43}$ \\
\hline
\end{tabular}
\begin{tablenotes}
\footnotesize
\item The radius values correspond to observed estimates of the compact objects.
\end{tablenotes}
\end{threeparttable}
\end{table}

\begin{figure}[bt]
\centering
\begin{minipage}{0.44\textwidth}
    \centering
    \includegraphics[width=\textwidth]{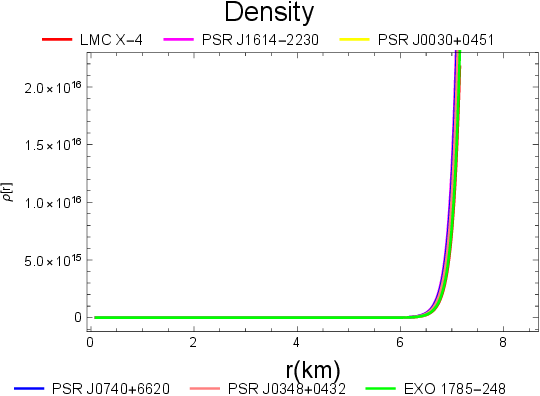}
    \\ \small (a)
\end{minipage}
\hfill
\begin{minipage}{0.44\textwidth}
    \centering
    \includegraphics[width=\textwidth]{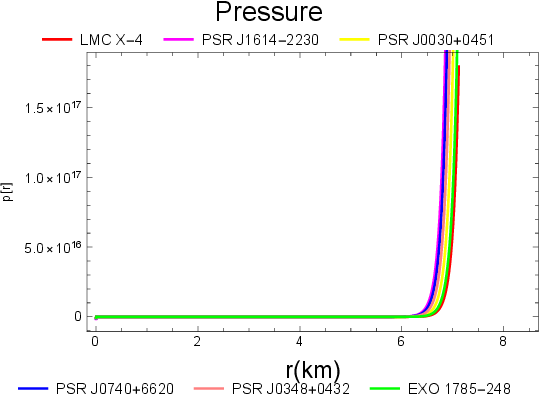}
    \\ \small (b)
\end{minipage}
\caption{(a) Evolution of density about radius $r$ for exponential model; (b) Evolution of pressure about radius $r$ for exponential model.}
\label{f2}
\end{figure}

In Fig. \ref{f2}, we have depicted the graphical demonstration of density and pressure \emph{LMC X-4, PSR J0740+6620, PSR J1614-2230, PSR J0348+0432, PSR J0030+0451, and EXO 1785-248}. The left panel figure illustrates that density exhibits a monotonic increasing pattern with respect to radius, reaching its maximum value at the surface. Hence, it can be concluded that there is enhanced density concentration near the outer region of the configuration. This indicates that the configurations exhibit a relatively larger density near the outer region compared to the inner region within the adopted framework, which also supports our previous results, as concluded after Fig. \ref{f1}. Due to a higher density shell, the pressure at the outer surface of the star also shows a monotonic increasing pattern in the right panel figure in Fig. \ref{f2}. 

\subsubsection*{Energy conditions}
To verify the existence of non-exotic matter in star formations, models must satisfy the energy constraints of general relativity. These conditions can be used to extensively analyze the Hawking-Penrose singularity theorems and the second rule of black hole thermodynamics \cite{hawking1976particle}. Energy circumstances have also been cited as a major explanation for a number of intriguing cosmic events \cite{santos2007energy,balart2014regular}. They play an important role in future singularity studies and in our knowledge of the dark energy (DE) phase. Furthermore, in order to effectively depict the unique behavior of matter in models of compact objects, these characteristics must be met. It is possible to consider the proposed stellar models to be physically possible when these conditions are satisfied \cite{das2024realistic}:

$(i)$ $NEC : {\rho } +{p} \geq 0 $

$(ii)$ $SEC : {\rho } + {p} \geq 0 $ , ${\rho } + 3 {p}  \geq 0$. \\
The inequalities above must be met by an compact objects for the energy-momentum tensor to maintain the positivity uniformly throughout the stellar configuration. Although some research has indicated that unphysical stress-energy tensors are the exclusive source of energy condition violations, this is not necessarily the case. Specifically, situations with minimally coupled scalar fields or theories with curvature-coupled scalar fields frequently fail the Strong Energy Condition (SEC). Therefore, it can be unduly idealistic to rely only on the SEC as a guiding principle. However, the Weak Energy Condition (WEC) and the Null Energy Condition (NEC) are not always violated in conjunction with a violation of the SEC. 

\begin{figure}[bt]
\centering
\begin{minipage}{0.44\textwidth}
    \centering
    \includegraphics[width=\textwidth]{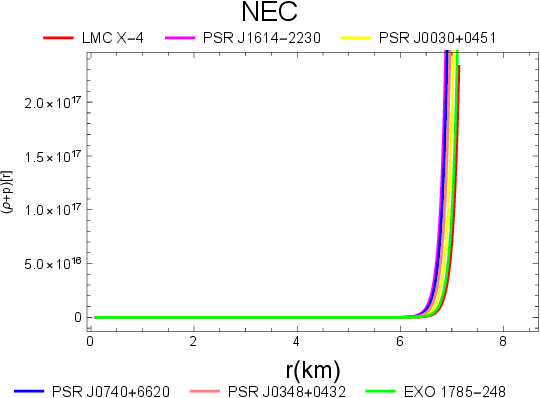}
    \\ \small (a)
\end{minipage}
\hfill
\begin{minipage}{0.44\textwidth}
    \centering
    \includegraphics[width=\textwidth]{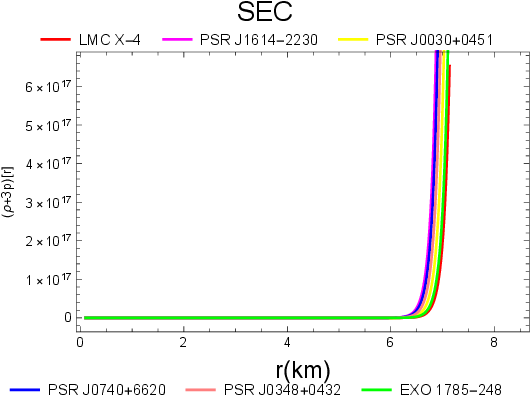}
    \\ \small (b)
\end{minipage}
\caption{(a) Evolution of Null Energy Condition about radius $r$ for exponential model; (b) Evolution of Strong Energy Condition about radius $r$ for exponential model.}
\label{f3}
\end{figure}
The SEC and NEC evolution have been demonstrated in Fig. \ref{f3}, which confirms that the energy conditions have been verified for our model. Here, we can observe that the NEC is demonstrated in the left panel figure and the SEC is verified in the right panel figure. Both the NEC and SEC show a monotonic increasing pattern with respect to radius. The increasing rate of NEC and SEC indicates that, as we move closer to the core, the gravitational source of the star increases, which requires a steeper pressure gradient to counteract gravity and more internal stress to maintain stability. Also, the matter distribution of the star is more resistant to compression and leads to a stable compact configuration.

\subsubsection*{Stability analysis}
The adiabatic index $\Gamma=\left(\frac{\rho +p}{p}\right)v^2$, which is the ratio of the two specific temperatures, indicates the rigidity of the equation of state for a given density and $v^2$ is the square speed of sound defined as $v^2  = d{p}/d{\rho }$. In the interior area, the adiabatic index $\Gamma$ should be greater than 4/3 \cite{heintzmann1975neutron}. This index is important for analyzing if the generation of the pulsars is in a balanced condition under an infinitesimal radial adiabatic effect. This limiting value was proposed by Chandrashekhar \cite{chandrasekhar1964dynamical} to assess the dynamical stability of the anisotropic realistic stars with spherical symmetry in its occurrence of minor perturbations in radial adiabatic.
Note that Bondi \cite{royal1869proceedings} asserts that the model is unstable for $\Gamma $ less than $4/3$ in this context. According to Heintzmann and Hillebrandt \cite{heintzmann1975neutron}, the relativistic stability requirement, $\Gamma $, should be bigger than 4/3 for compact objects with an increasing anisotropy factor. Consequently, \cite{royal1869proceedings} and \cite{heintzmann1975neutron} align with the previously presented model. 

\begin{figure}[bt]
\centering
\begin{minipage}{0.64\textwidth}
    \centering
    \includegraphics[width=\textwidth]{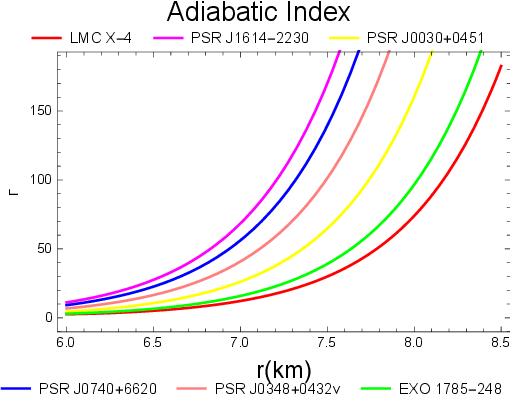}
    \\ \small (a)
\end{minipage}
\caption{(a) Evolution of Adiabatic Index about radius $r$ for exponential model.}
\label{f4}
\end{figure}

Fig. \ref{f4} demonstrates the adiabatic index graphically for \emph{LMC X-4, PSR J0740+6620, PSR J1614-2230, PSR J0348+0432, PSR J0030+0451, and EXO 1785-248} using different color grades. The graph displays a monotonic increase in pattern within the interior of the stars, and the graphical depiction verifies that the adiabatic index for all compact stellar configurations fits within the permitted limit, which is $ > 4/3$. Increasing $\Gamma$ indicates the material moves into a phase where nuclear forces strongly oppose additional compression as density increases, supporting compact stellar configurations up to their maximum stable mass. An increasing adiabatic index indicates that the system is generally more resistant to gravitational collapse, the effective matter distribution becomes more resistant to compression, and pressure increases more strongly with density. It indicates that nuclear matter is more difficult to compress at higher densities, which improves stability in compact stellar configurations.

\subsection*{Mass function as a quadratic form of radius}

In this section, we consider the bosonic mass as a quadratic function of the radial coordinate
\begin{equation}
m_B=x-yr+zr^2,
    \label{E23}
\end{equation}
where $x$, $y$, and $z$ are arbitrary constants. Hence, the modified mass for compact stars is
\begin{equation}
M = \frac{2}{5} \sqrt{2} \, \sqrt{1 + r^{2}} \,  \left( x + r \left( -y + r z \right) \right)^{4} \, \alpha \, \left( -1 + (1 + r^{2})^{3/4} \, \alpha^{3/2} \right) ,
 \label{E24}
\end{equation}
and compactness is of the form
\begin{equation}
   u= \frac{2 \sqrt{2} \, \sqrt{1 + r^2} \, \left(x + r(-y + rz)\right)^4 \alpha \left(-1 + (1 + r^2)^{\tfrac{3}{4}} \alpha^{\tfrac{3}{2}}\right)}{5r} ,
\label{E34}
\end{equation}
and the surface redshift is
\begin{equation}
    z=-1 + \frac{1}{\sqrt{1 - \frac{4 \sqrt{2} \, \sqrt{1 + r^2} \, (x + r(-y + rz))^4 \, \alpha \left(-1 + (1 + r^2)^{\tfrac{3}{4}} \alpha^{\tfrac{3}{2}}\right)}{5r}}}
\label{E35}
\end{equation}
The numerical values for the mass of the compact astrophysical configurations and the compactness have been represented in Table \ref{T3}, to understand the nature of the compact objects in the quadratic model.  

\begin{table}[bt]
\centering
\caption{From Eq.~(\ref{E24}), the model mass, compactness, and surface redshift for specific pulsars have been evaluated for the quadratic model}
\label{T3}
\begin{threeparttable}
\begin{tabular}{|c|c|c|c|c|c|c|}
\hline
\textbf{Sl. No.} & \textbf{Star Model} & 
\textbf{Mass ($M/M_{\odot}$)} & 
$\boldsymbol{\alpha}$ & 
\textbf{Model Mass ($M_{\odot}$)} & 
\textbf{Compactness} & 
\textbf{Surface Redshift} \\
\hline

1 & LMC X-4 & $1.04 \pm 0.09$ & 1.20 & 2.00526 & 0.250658 & 0.416079 \\
\hline

2 & PSR J0740+6620 & $2.07 \pm 0.11$ & 1.45 & 1.99377 & 0.162095 & 0.216432 \\
\hline

3 & PSR J1614-2230 & $1.908 \pm 0.016$ & 1.50 & 2.68384 & 0.206449 & 0.305099 \\
\hline

4 & PSR J0348+0432 & $2.01 \pm 0.04$ & 1.38 & 2.22105 & 0.170850 & 0.232503 \\
\hline

5 & PSR J0030+0451 & $1.44 \pm 0.16$ & 1.30 & 1.62243 & 0.129794 & 0.162154 \\
\hline

6 & EXO 1785-248 & $1.30 \pm 0.20$ & 1.23 & 1.27633 & 0.126120 & 0.156429 \\
\hline
\end{tabular}
\begin{tablenotes}
\footnotesize
\item $M$ denotes the estimated observed mass of the corresponding compact objects.
\end{tablenotes}
\end{threeparttable}
\end{table}

\begin{figure}[bt]
\centering
\begin{minipage}{0.44\textwidth}
    \centering
    \includegraphics[width=\textwidth]{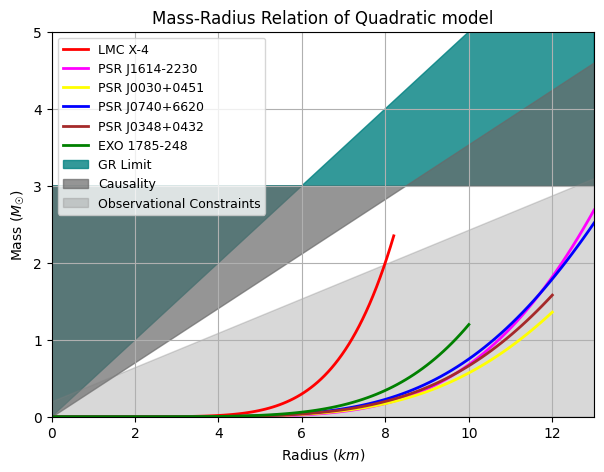}
    \\ \small (a)
\end{minipage}
\hfill
\begin{minipage}{0.44\textwidth}
    \centering
    \includegraphics[width=\textwidth]{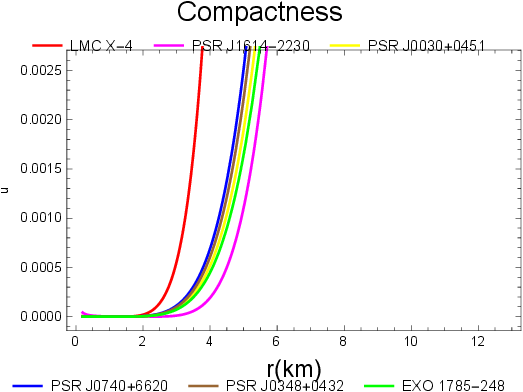}
    \\ \small (b)
\end{minipage}
\hfill
\begin{minipage}{0.44\textwidth}
    \centering
    \includegraphics[width=\textwidth]{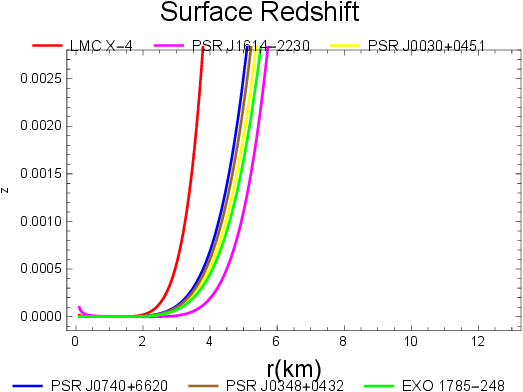}
    \\ \small (c)
\end{minipage}
\caption{(a) Evolution of mass with radius $r$ for the quadratic model; (b) Evolution of compactness with radius $r$ for the quadratic model; (c) Evolution of surface redshift with radius $r$ for the quadratic model.}
\label{f5}
\end{figure}
The mass, compactness, and surface redshift are graphically depicted in Fig. \ref{f5}. The mass evolution in relation to the radius is shown in the figure on the left side. Here we observe the mass distribution for the stellar models \emph{LMC X-4, PSR J0740+6620, PSR J1614-2230, PSR J0348+0432, PSR J0030+0451, and EXO 1785-248}, which show a monotonic increase in mass. It attains the maximum mass at the boundary, as the graphical representation shows the accumulated mass of the whole interior star. It is noted that for all the stellar models, the mass lies within the observed mass limit for the respective models. Also, the right panel figure shows the compactness and this has an increasing pattern. The numerical value of the compactness in Table \ref{T1} confirms that these lie within Buchdahl's  limit, and the obtained compactness values fall within the range commonly associated with compact stellar configurations. The surface redshift evaluation shows how much light energy is lost or redshifted when you escape from the gravitational field. In our model, we observe that the surface redshift rises with increasing radius. Because of the greater mass value, we can conclude that although the gravitational pull is still present at core pressure, the gravitational potential is stronger, and the rate of photon particle emission from the outer layer is lower near the surface. The light energy loses the most energy at the boundary. The lower panel figure demonstrates the graphical representation of the surface redshift, which has an increasing pattern confirming that the photons lose more energy while escaping from the shell of the star.

\subsubsection*{Physical attributes}
Using Eq. (\ref{E04}) we have reformulated the density of the stars for quadratic bosonic mass function as follows
\begin{equation}
\begin{split}
\rho &= 
\frac{(x + r(-y + r z))^3 \, \alpha}{5 \sqrt{2}\, \pi^2\, r^3 \sqrt{1 + r^2}} \\
&\quad \times 
\Biggl[
r^2 y \left( 10 - 13 (1 + r^2)^{3/4} \alpha^{3/2} \right)
- 8 y \left( -1 + (1 + r^2)^{3/4} \alpha^{3/2} \right) \\
&\qquad + 3 r^3 z \left( -6 + 7 (1 + r^2)^{3/4} \alpha^{3/2} \right)
+ r \Bigl(
16 z \left( -1 + (1 + r^2)^{3/4} \alpha^{3/2} \right)
+ x \left( -2 + 5 (1 + r^2)^{3/4} \alpha^{3/2} \right)
\Bigr)
\Biggr]
\end{split}
\label{E25}
\end{equation}
  Hence, the number of particles per unit volume using Eq. (\ref{E07}) for the quadratic bosonic mass function is
  \begin{equation}
n= \frac{(x + r(-y + r z))^4 \, (-1 + \alpha^2)^2}{ 32 \pi^4} ,
\label{E26}
  \end{equation}
and the reformulated pressure using Eq. (\ref{E21}) is
\begin{equation}
\begin{split}
p &= 
\frac{(x + r(-y + r z))^3 \, \alpha}{160 \pi^4}
\Biggl[
5 (x + r(-y + r z))^2 (-1 + \alpha^2)^2 \\
&\quad
- \frac{16 \sqrt{2} \pi^2}{r^3 \sqrt{1 + r^2}}
\Bigl(
8y - 8 (1 + r^2)^{3/4} y \alpha^{3/2}
+ r^2 y \left( 10 - 13 (1 + r^2)^{3/4} \alpha^{3/2} \right) \\
&\qquad
+ 3 r^3 z \left( -6 + 7 (1 + r^2)^{3/4} \alpha^{3/2} \right)
+ r \bigl(
16 z \left( -1 + (1 + r^2)^{3/4} \alpha^{3/2} \right)
+ x \left( -2 + 5 (1 + r^2)^{3/4} \alpha^{3/2} \right)
\bigr)
\Bigr)
\Biggr]
\end{split}
\label{E27}
\end{equation}

\begin{table}[bt]
\centering
\caption{Evaluation of density, pressure, and number of particles per unit volume for the quadratic model of bosonic mass function at the surface}
\label{T4}
\begin{threeparttable}
\begin{tabular}{|c|c|c|c|c|c|}
\hline
\textbf{Sl. No.} & \textbf{Star Model} & 
\textbf{Radius (km)} & 
\textbf{Density} & 
\textbf{Pressure} & 
\textbf{Number of Particles/Volume} \\
& & & $(\mathrm{MeV/fm^3})$ & $(\mathrm{MeV/fm^3})$ & \\
\hline

1 & LMC X-4 & $8.301 \pm 0.2$ 
& $3.68138 \times 10^{8}$ 
& $1.73387 \times 10^{9}$ 
& $3.57596 \times 10^{6}$ \\
\hline

2 & PSR J0740+6620 & $12.34^{+1.89}_{-1.67}$ 
& $2.7621 \times 10^{8}$ 
& $1.74055 \times 10^{10}$ 
& $2.43852 \times 10^{7}$ \\
\hline

3 & PSR J1614-2230 & $13 \pm 2$ 
& $5.15939 \times 10^{7}$ 
& $1.9381 \times 10^{10}$ 
& $2.69132 \times 10^{7}$ \\
\hline

4 & PSR J0348+0432 & $13 \pm 2$ 
& $4.20298 \times 10^{7}$ 
& $9.31658 \times 10^{9}$ 
& $1.40884 \times 10^{7}$ \\
\hline

5 & PSR J0030+0451 & $13.02 \pm 1.24$ 
& $2.10064 \times 10^{8}$ 
& $5.99978 \times 10^{9}$ 
& $9.55212 \times 10^{6}$ \\
\hline

6 & EXO 1785-248 & $10.10 \pm 0.44$ 
& $2.15944 \times 10^{8}$ 
& $3.05998 \times 10^{9}$ 
& $5.31626 \times 10^{6}$ \\
\hline
\end{tabular}
\begin{tablenotes}
\footnotesize
\item The radius values correspond to the observed estimates of the compact objects.
\end{tablenotes}
\end{threeparttable}
\end{table}

\begin{figure}[bt]
\centering
\begin{minipage}{0.44\textwidth}
    \centering
    \includegraphics[width=\textwidth]{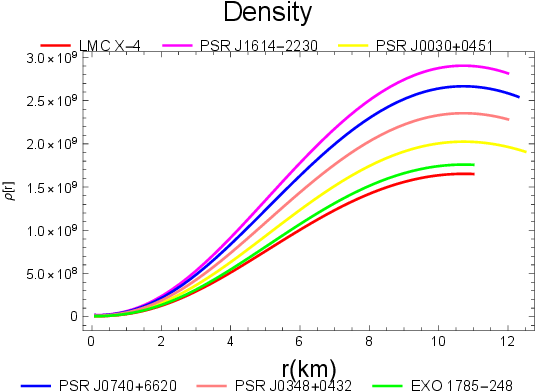}
    \\ \small (a)
\end{minipage}
\hfill
\begin{minipage}{0.44\textwidth}
    \centering
    \includegraphics[width=\textwidth]{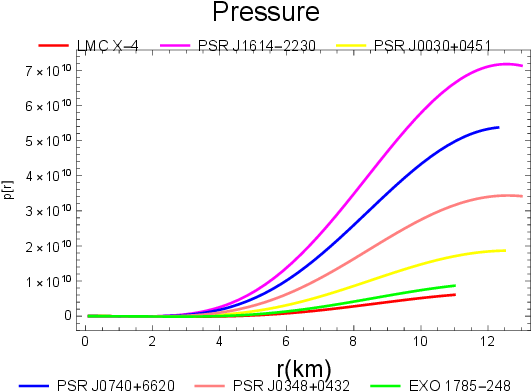}
    \\ \small (b)
\end{minipage}
\caption{(a) Evolution of density about radius $r$ for the quadratic model; (b) Evolution of pressure about radius $r$ for the quadratic model.}
\label{f6}
\end{figure}
In Fig. \ref{f6}, the left panel figure represents the density, and the right panel figure represents the pressure for the star models in case of the quadratic model. The graphical representations demonstrate the star model \emph{LMC X-4, PSR J0740+6620, PSR J1614-2230, PSR J0348+0432, PSR J0030+0451, and EXO 1785-248} using different color grades. Here, we can observe that the left panel figure illustrates that density exhibits a monotonic increasing pattern with respect to radius, reaching its maximum value near 10 km radius and then showing asymptotic flatness to the surface for the model \emph{ PSR J0740+6620, PSR J1614-2230, PSR J0348+0432, and PSR J0030+0451}, whereas for \emph{LMC X-4, and EXO 1785-248} maximum density is near about 8.5 km radius and then show asymptotic flatness. From this graphical representation, it can be concluded that there is enhanced density concentration near the outer region of the configuration. This indicates that the configurations exhibit a relatively larger density near the outer region compared to the inner region within the adopted framework. The two primary sections of compact stellar configurations are the core, which is composed of liquid nuclear matter, and the crust, which is composed of crystalline lattices of neutron-rich nuclei. Because of the greater chemical potential of electrons, which encourages electron capture, nuclei in the deeper region of the compact stellar configuration crust are more neutron-rich \cite{Watanabe2011nuclear}. Rapid rotation causes the mass to be redistributed during accretion, which results in the outer region exhibiting larger effective density accumulation. Additionally, for each star model, the density accumulation achieves its maximum level; this is indicated by the asymptotic flatness of density at the star's surface. The mass buildup and increasing density pattern are supported by the increasing pressure evolution with respect to radius, which is shown in the right panel figure. The pressure is therefore sufficiently high to regulate the bulk accumulation. Therefore, the pressure is adequate to keep the balance. The work of \cite{das2023study,Vogt2010newtonian} also shows an increasing density trend in case of study of compact stellar configurations.

\subsubsection*{Energy conditions}

\begin{figure}[bt]
\centering
\begin{minipage}{0.44\textwidth}
    \centering
    \includegraphics[width=\textwidth]{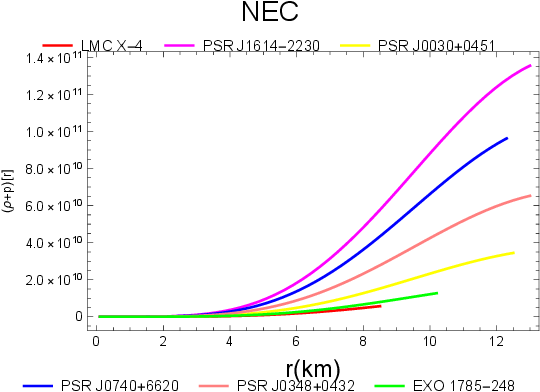}
    \\ \small (a)
\end{minipage}
\hfill
\begin{minipage}{0.44\textwidth}
    \centering
    \includegraphics[width=\textwidth]{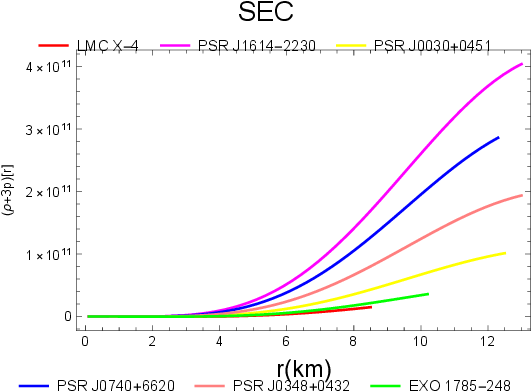}
    \\ \small (b)
\end{minipage}
\caption{(a) Evolution of Null Energy Condition about radius $r$ for quadratic model; (b) Evolution of Strong Energy Condition about radius $r$ for quadratic model.}
\label{f7}
\end{figure}

Fig. \ref{f7} shows the evolution of the SEC and NEC, confirming that the energy conditions for our model have been confirmed. Here, we can see that the right panel graphic verifies the SEC, whereas the left panel figure illustrates the NEC. In relation to radius, the NEC and SEC both exhibit a monotonically increasing pattern. The rising NEC and SEC rates show that the gravitational source of the star rises with proximity to the core, leading to a steeper pressure gradient to oppose gravity and more internal stress to preserve stability. Additionally, a stable compact configuration results from mass of the star distribution being more resistant to compression.

\subsubsection*{Stability analysis}

\begin{figure}[bt]
\centering
\begin{minipage}{0.64\textwidth}
    \centering
    \includegraphics[width=\textwidth]{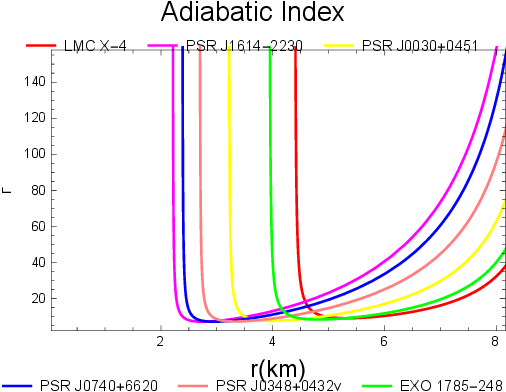}
    \\ \small (a)
\end{minipage}
\caption{(a) Evolution of Adiabatic Index about radius $r$ for quadratic model.}
\label{f8}
\end{figure}

Adiabatic index is important in understanding how the change in pressure is affected by the change in density. A higher value of the adiabatic index indicates pressure increases rapidly with the increase of density, hence the star model is more resistant to compression and leads to a more stable formation. In our study, Fig. \ref{f8} demonstrates the graphical representation of adiabatic index with respect to radius for the star model \emph{LMC X-4, PSR J0740+6620, PSR J1614-2230, PSR J0348+0432, PSR J0030+0451, and EXO 1785-248}. Here, we can see that the adiabatic index reaches its maximum value close to the core region, which is approximately 4.5 km from the center. This suggests that the region is distinguished by increased effective pressure support and effective pressure contribution, which makes matter more resistant to compression and forms a stable core. In the intermediate region, from 4.5 km to 6 km region it has a decaying pattern and from 6 km to the surface of the stars, the adiabatic index again increases. However, the decaying value of the adiabatic index does not cross the Heintzmann and Hillebrandt \cite{heintzmann1975neutron} stability requirement, which is $\Gamma $, should be bigger than 4/3. Hence, we can conclude that our model is well-stable within the entire region of the star. Anyway, near the surface, the increasing pattern of adiabatic index ensures that effective pressure contribution is dominating and matter is resistant to compression.  

We would like to reiterate that the observational compact-object data presented here are used only as representative astrophysical scales for comparison of mass and compactness, and not as direct evidence for AdS bosonic stars.

\subsection*{Mass function as power law form of radius}

In this section, we consider the bosonic mass as power law form of the radial coordinate
\begin{equation}
m_B= \xi r^\zeta ,
    \label{E28}
\end{equation}
where $\xi$ and $\zeta$ are arbitrary constants. Hence, the modified mass for the compact stars evaluated as
\begin{equation}
M=\frac{2}{5} \sqrt{2}\, r^{4\zeta} \sqrt{1 + r^{2}}\, \xi^{4} \alpha \left( -1 + (1 + r^{2})^{\frac{3}{4}} \alpha^{\frac{3}{2}} \right),
    \label{E29}
\end{equation}
and the compactness is
\begin{equation}
u=M/r=\frac{2}{5} \sqrt{2} \, r^{-1 + 4\zeta} \sqrt{1 + r^2} \, \xi^4 \, \alpha \left(-1 + (1 + r^2)^{\tfrac{3}{4}} \alpha^{\tfrac{3}{2}}\right),
    \label{E30}
\end{equation}
and the surface redshift is
\begin{equation}
z=(1-2u)^{-1/2} -1=-1 + \frac{1}{\sqrt{1 - \frac{4}{5} \sqrt{2} \, r^{-1 + 4\zeta} \sqrt{1 + r^2} \, \xi^4 \, \alpha \left(-1 + (1 + r^2)^{\tfrac{3}{4}} \alpha^{\tfrac{3}{2}}\right)}}.
    \label{E31}
\end{equation}

To comprehend the nature of the compact items in the power law model, Table \ref{T5} presents the numerical values for the mass of the compact astrophysical configurations and the compactness.  

\begin{table}[bt]
\centering
\caption{From Eq.(\ref{E29}), the model mass, compactness, and surface redshift for specific pulsars have been evaluated for the power law model}
\label{T5}
\begin{threeparttable}
\begin{tabular}{|c|c|c|c|c|c|c|}
\hline
\textbf{Sl. No.} & \textbf{Star Model} & 
\textbf{Mass ($M/M_{\odot}$)} & 
$\boldsymbol{\alpha}$ & 
\textbf{Model Mass ($M_{\odot}$)} & 
\textbf{Compactness} & 
\textbf{Surface Redshift} \\
\hline

1 & LMC X-4 & $1.04 \pm 0.09$ & 1.20 & 2.00356 & 0.235713 & 0.375457 \\
\hline

2 & PSR J0740+6620 & $2.07 \pm 0.11$ & 1.45 & 2.37338 & 0.192958 & 0.276104 \\
\hline

3 & PSR J1614-2230 & $1.908 \pm 0.016$ & 1.50 & 2.99798 & 0.230614 & 0.362378 \\
\hline

4 & PSR J0348+0432 & $2.01 \pm 0.04$ & 1.38 & 1.30201 & 0.100155 & 0.118250 \\
\hline

5 & PSR J0030+0451 & $1.44 \pm 0.16$ & 1.30 & 2.48437 & 0.198750 & 0.288313 \\
\hline

6 & EXO 1785-248 & $1.30 \pm 0.20$ & 1.23 & 1.78337 & 0.176222 & 0.242686 \\
\hline
\end{tabular}
\begin{tablenotes}
\footnotesize
\item $M$ denotes the estimated observed mass of the corresponding compact objects.
\end{tablenotes}
\end{threeparttable}
\end{table}

\begin{figure}[bt]
\centering
\begin{minipage}{0.44\textwidth}
    \centering
    \includegraphics[width=\textwidth]{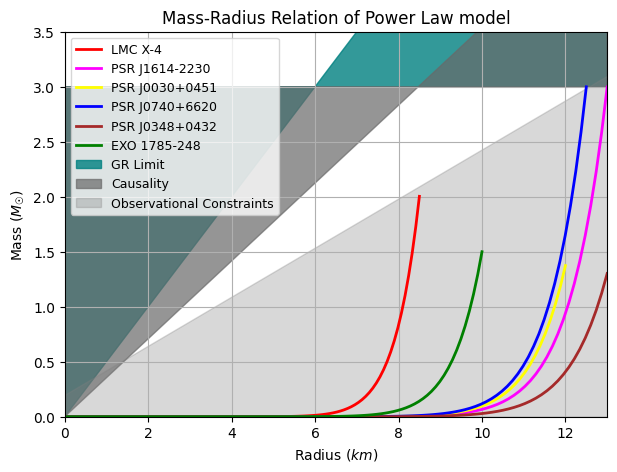}
    \\ \small (a)
\end{minipage}
\hfill
\begin{minipage}{0.44\textwidth}
    \centering
    \includegraphics[width=\textwidth]{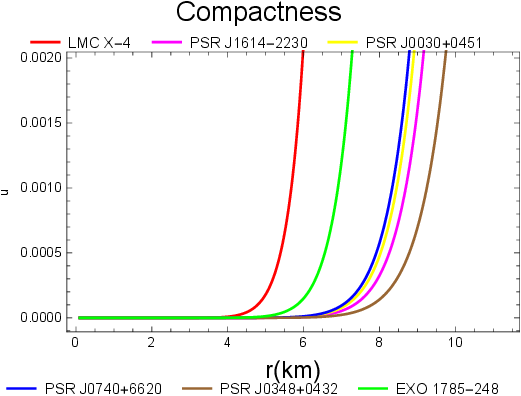}
    \\ \small (b)
\end{minipage}
\hfill
\begin{minipage}{0.44\textwidth}
    \centering
    \includegraphics[width=\textwidth]{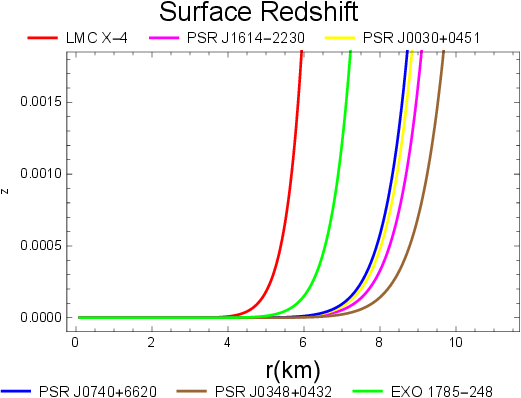}
    \\ \small (c)
\end{minipage}
\caption{(a) Evolution of mass about radius $r$ for the power law model; (b) Evolution of compactness about radius $r$ for the power law model; (c) Evolution of surface redshift about radius $r$ for the power law model.}
\label{f9}
\end{figure}

The mass, compactness, and surface redshift are graphically represented in Fig. \ref{f9}. The mass evolution is demonstrated in the left panel figure for the star models \emph{LMC X-4, PSR J0740+6620, PSR J1614-2230, PSR J0348+0432, PSR J0030+0451, and EXO 1785-248} with the help of different color grades. We can observe \emph{PSR J0740+6620, PSR J1614-2230} attains maximum mass compared to other star models. Also, the mass shows monotonic increasing pattern and has maximum value at the boundary. Compactness is shown in the right panel, and the surface redshift is shown in the lower panel. Our model is stable and the star models are compact stellar configurations, as confirmed by the compactness in the graphic representation and numerical values in Table \ref{T5}, which guarantee that the compactness lies within Buchdahl's limit \cite{das2025compact}. The amount that light energy is redshifted or lost during escape from the gravitational field is indicated by the surface redshift evaluation. We find that the surface redshift increases as the radius increases in our model. Thus, we may deduce that while the gravitational pull persists at core pressure, the gravitational potential is stronger and the rate of photon particle emission from the outer layer is lower toward the surface because of the higher mass value. At the boundary, the light energy has the greatest energy loss. Also, the numerical value of the gravitational redshift lies within the permissible limit within the stellar models, which is $\leq 5.211$ \cite{das2025evolution} as confirmed in Table \ref{T5}. 

\subsubsection*{Physical attributes}

Using Eq. (\ref{E04}) we have reformulated the density of the stars for the power law form of bosonic mass function as follows
\begin{equation}
\begin{array}{cc}
    \rho= \frac{r^{-4 + 4\zeta} \xi^4 \alpha \left[ 8\zeta \left(-1 + (1 + r^2)^{\frac{3}{4}} \alpha^{\frac{3}{2}}\right) + r^2 \left(-2 + 5(1 + r^2)^{\frac{3}{4}} \alpha^{\frac{3}{2}} + 8\zeta \left(-1 + (1 + r^2)^{\frac{3}{4}} \alpha^{\frac{3}{2}}\right) \right) \right]}{5 \sqrt{2} \pi^2 \sqrt{1 + r^2}}
\label{E36}
\end{array}
\end{equation}
  Hence, the number of particles per unit volume using Eq. (\ref{E07}) for the quadratic bosonic mass function is
  \begin{equation}
n= \frac{r^{4\zeta} \xi^4 \left(-1 + \alpha^2\right)^2}{32 \pi^4},
\label{E37}
  \end{equation}
and the reformulated pressure using Eq. (\ref{E21}) is
\begin{equation}
\begin{array}{cc}
    p=\frac{r^{4\zeta} \xi ^4 \alpha \left[ 5 r^\zeta \xi \left(-1 + \alpha^2\right)^2 - \frac{ 16 \sqrt{2} \pi^2 \left(8 \zeta \left(-1 + (1 + r^2)^{\frac{3}{4}} \alpha^{\frac{3}{2}}\right)+ r^2 \left(-2 - 8\zeta + 5(1 + r^2)^{\frac{3}{4}} \alpha^{\frac{3}{2}} + 8(1 + r^2)^{\frac{3}{4}} \zeta \alpha^{\frac{3}{2}}\right)\right)}{r^4 \sqrt{1 + r^2}}\right]}{160 \pi^4}.
\label{E38}
\end{array}
\end{equation}

\begin{table}[bt]
\centering
\caption{Evaluation of density, pressure, and number of particles per unit volume for the power law form of bosonic mass function at the surface}
\label{T6}
\begin{threeparttable}
\begin{tabular}{|c|c|c|c|c|c|}
\hline
\textbf{Sl. No.} & \textbf{Star Model} & 
\textbf{Radius (km)} & 
\textbf{Density} & 
\textbf{Pressure} & 
\textbf{Number of Particles/Volume} \\
& & & $(\mathrm{MeV/fm^3})$ & $(\mathrm{MeV/fm^3})$ & \\
\hline
1 & LMC X-4 & $8.301 \pm 0.2$ 
& $9.0845 \times 10^{8}$ 
& $4.6872 \times 10^{8}$ 
& $2.54963 \times 10^{6}$ \\
\hline

2 & PSR J0740+6620 & $12.34^{+1.89}_{-1.67}$ 
& $1.36265 \times 10^{8}$ 
& $8.27479 \times 10^{8}$ 
& $2.37835 \times 10^{6}$ \\
\hline

3 & PSR J1614-2230 & $13 \pm 2$ 
& $5.1083 \times 10^{7}$ 
& $2.62918 \times 10^{8}$ 
& $9.92398 \times 10^{5}$ \\
\hline

4 & PSR J0348+0432 & $13 \pm 2$ 
& $2.21868 \times 10^{7}$ 
& $4.71335 \times 10^{7}$ 
& $2.78343 \times 10^{5}$ \\
\hline

5 & PSR J0030+0451 & $13.02 \pm 1.24$ 
& $1.12926 \times 10^{8}$ 
& $2.64437 \times 10^{8}$ 
& $1.01635 \times 10^{6}$ \\
\hline

6 & EXO 1785-248 & $10.10 \pm 0.44$ 
& $1.15048 \times 10^{9}$ 
& $2.82301 \times 10^{9}$ 
& $6.20407 \times 10^{6}$ \\
\hline
\end{tabular}
\begin{tablenotes}
\footnotesize
\item The radius values correspond to the observed estimates of the compact objects.
\end{tablenotes}
\end{threeparttable}
\end{table}

\begin{figure}[bt]
\centering
\begin{minipage}{0.3\textwidth}
    \centering
    \includegraphics[width=\textwidth]{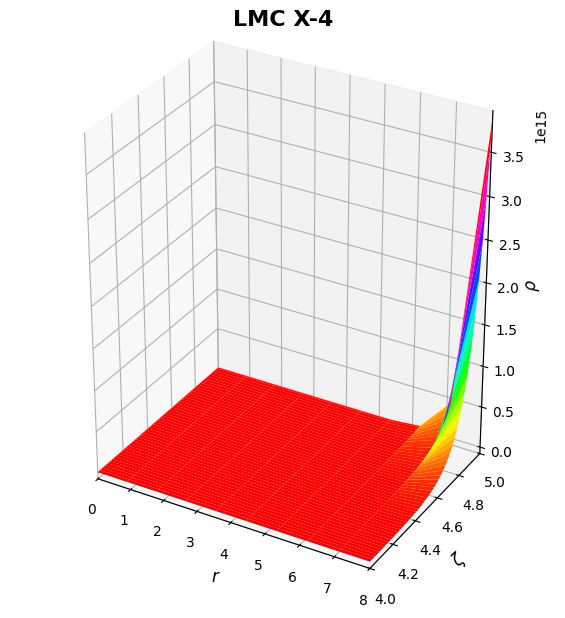}
    \\ \small (a)
\end{minipage}
\hfill
\begin{minipage}{0.3\textwidth}
    \centering
    \includegraphics[width=\textwidth]{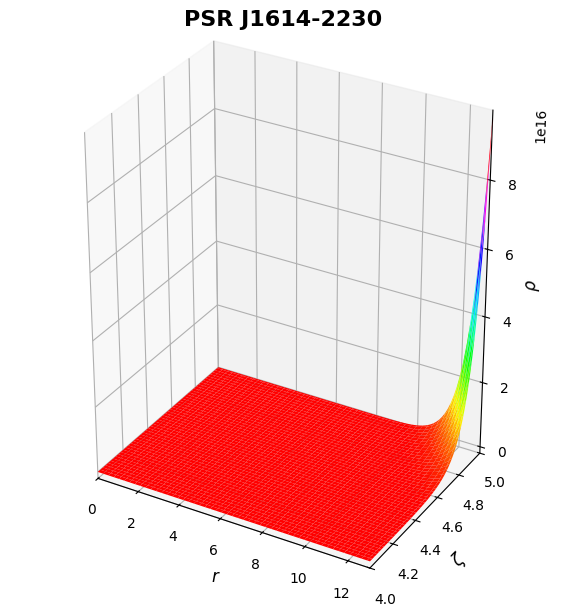}
    \\ \small (b)
\end{minipage}
\hfill
\begin{minipage}{0.3\textwidth}
    \centering
    \includegraphics[width=\textwidth]{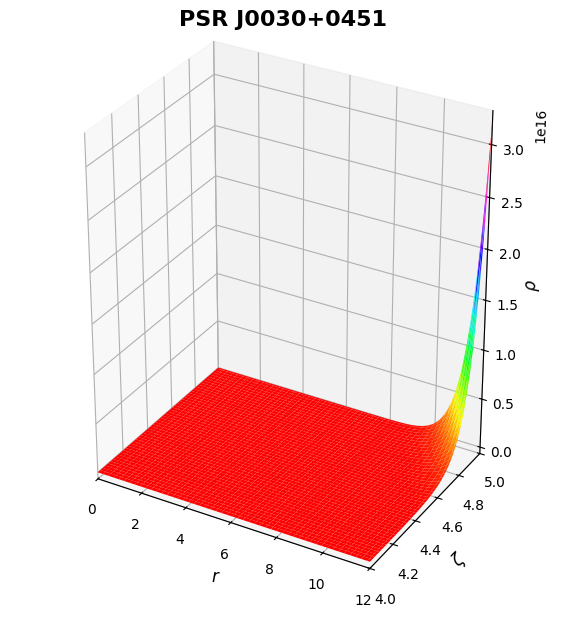}
    \\ \small (c)
\end{minipage}
\caption{ Evolution of density about radius $r$ for the power law model for the stellar models LMC X-4, PSR J1614-2230, and PSR J0030+0451. }
\label{f10}
\end{figure}

\begin{figure}[bt]
\centering
\begin{minipage}{0.3\textwidth}
    \centering
    \includegraphics[width=\textwidth]{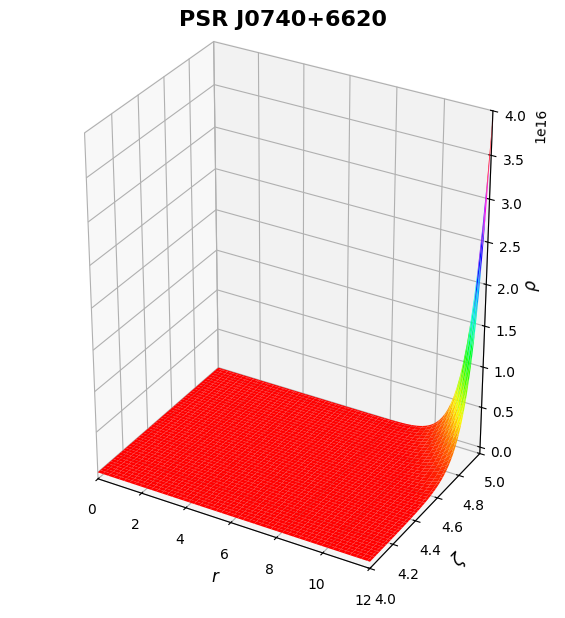}
    \\ \small (a)
\end{minipage}
\hfill
\begin{minipage}{0.3\textwidth}
    \centering
    \includegraphics[width=\textwidth]{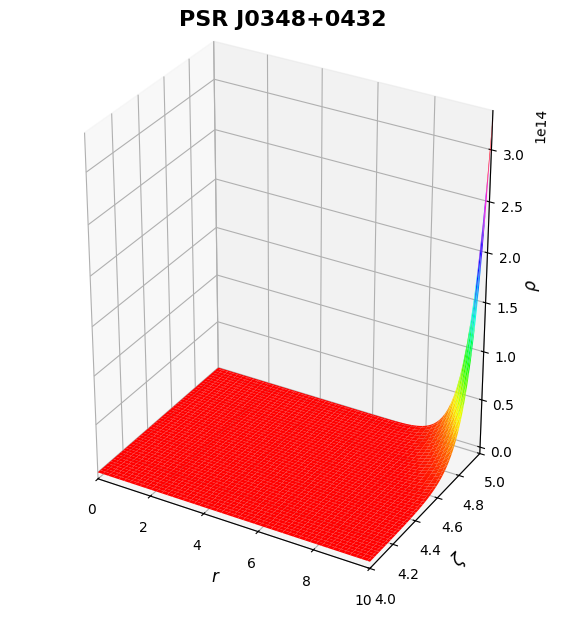}
    \\ \small (b)
\end{minipage}
\hfill
\begin{minipage}{0.3\textwidth}
    \centering
    \includegraphics[width=\textwidth]{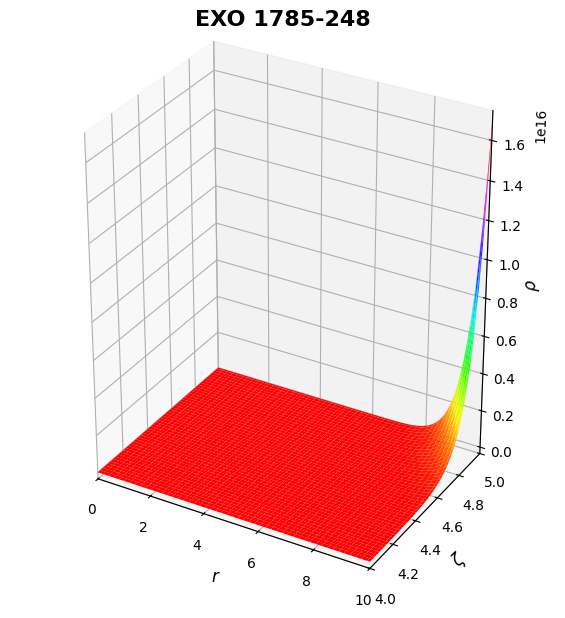}
    \\ \small (c)
\end{minipage}
\caption{ Evolution of density about radius $r$ for the power law model for the stellar models PSR J0348+0432, PSR J0348+0432 and EXO 1785-248. }
\label{f11}
\end{figure}

In Figs. \ref{f10} and \ref{f11}, we illustrated the density evolution with respect to radius and $\zeta$, the exponent of the power law form, to demonstrate how the density changes as the exponent in the power law (i.e., $\zeta$) and the radius increases. In Fig. \ref{f10} we have demonstrated the density evolution for the stellar objects \emph{LMC X-4, PSR J0740+6620, and PSR J1614-2230} and in Fig. \ref{f11} demonstrate the density evolution for the stellar objects \emph{PSR J0348+0432, PSR J0030+0451, and EXO 1785-248}. Since density increases with increasing exponent values, it can be concluded that there is an accretion around the star as we move toward the surface because there is a greater accumulation of matter particles than in the core. This also indicates that the stellar objects has denser shell compared to less denser core. The graphical demonstration also indicate that, for higher exponential value in the power law form the rate of accretion is higher, i.e. for $\zeta \geq 5$ the rate of escalation of of density also increases.  

\begin{figure}[bt]
\centering
\begin{minipage}{0.3\textwidth}
    \centering
    \includegraphics[width=\textwidth]{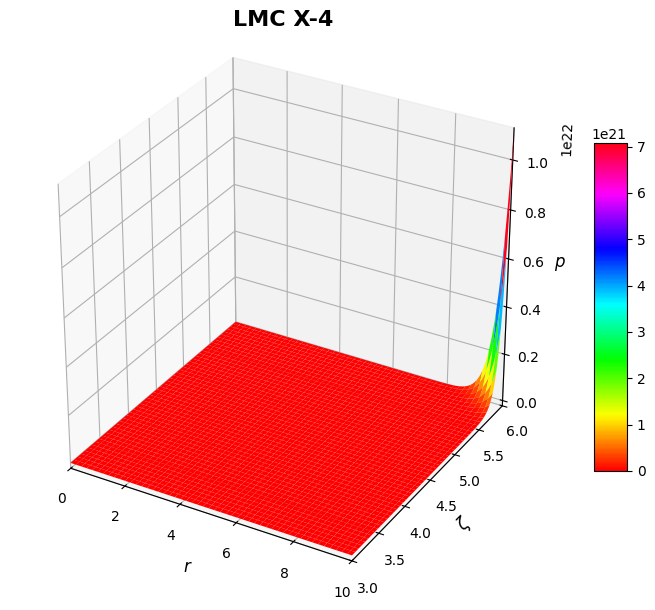}
    \\ \small (a)
\end{minipage}
\hfill
\begin{minipage}{0.3\textwidth}
    \centering
    \includegraphics[width=\textwidth]{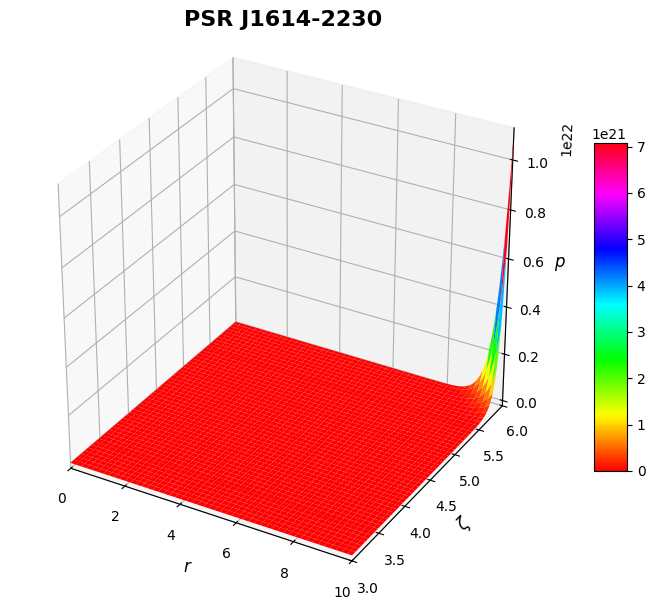}
    \\ \small (b)
\end{minipage}
\hfill
\begin{minipage}{0.3\textwidth}
    \centering
    \includegraphics[width=\textwidth]{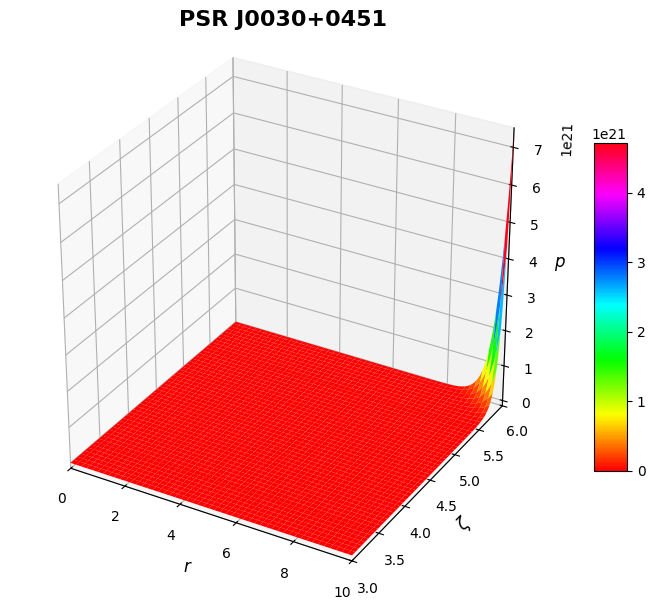}
    \\ \small (c)
\end{minipage}
\caption{ Evolution of the pressure about radius $r$ for the power law model for the stellar models LMC X-4, PSR J1614-2230, and PSR J0030+0451. }
\label{f12}
\end{figure}

\begin{figure}[bt]
\centering
\begin{minipage}{0.3\textwidth}
    \centering
    \includegraphics[width=\textwidth]{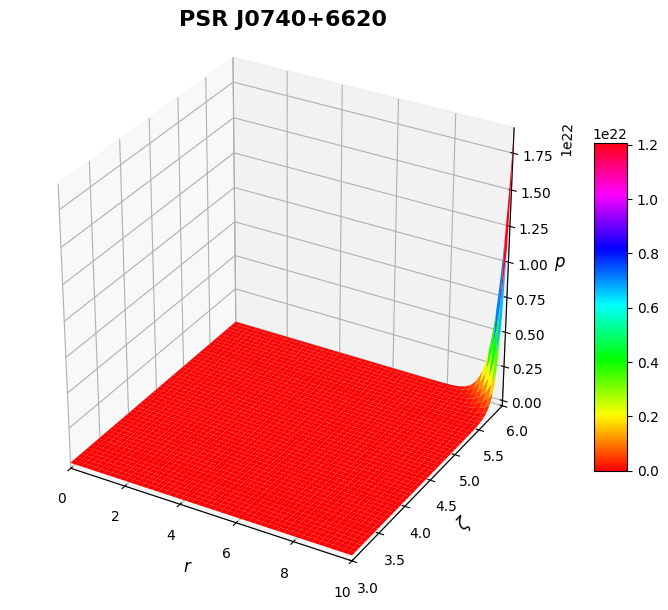}
    \\ \small (a)
\end{minipage}
\hfill
\begin{minipage}{0.3\textwidth}
    \centering
    \includegraphics[width=\textwidth]{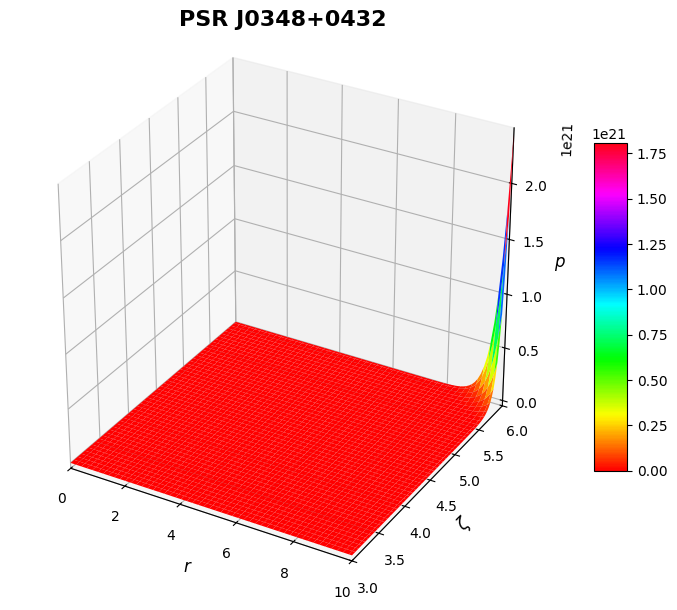}
    \\ \small (b)
\end{minipage}
\hfill
\begin{minipage}{0.3\textwidth}
    \centering
    \includegraphics[width=\textwidth]{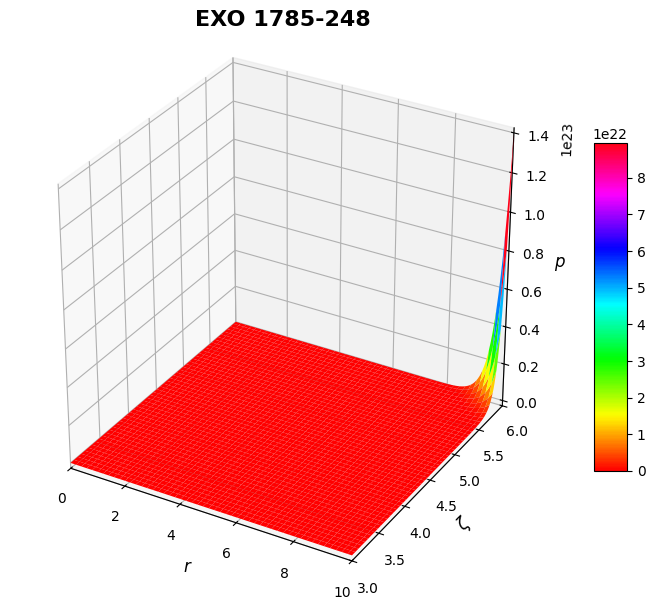}
    \\ \small (c)
\end{minipage}
\caption{ Evolution of the pressure about radius $r$ for the power law model for the stellar models PSR J0348+0432, PSR J0348+0432 and EXO 1785-248. }
\label{f13}
\end{figure}

We showed the evolution of pressure with respect to radius and $\zeta$, the exponent of the power law form, in Figs. \ref{f12} and \ref{f13} to show how the pressure evolves as the exponent in the power law and the radius increase. The pressure evolution for the stellar objects \emph{PSR J0348+0432, PSR J0030+0451, and EXO 1785-248} is shown in Fig. \ref{f12}, whereas the pressure evolution for the stellar objects \emph{LMC X-4, PSR J0740+6620, and PSR J1614-2230} is shown in Fig. \ref{f13}. As the radius increases—that is, as we approach the surface and for higher values of the exponent—we can see a monotonically increasing pattern in the pressure evolution. This finding validates the density evolution shown in Figs. \ref{f10} and \ref{f11}, since the evolution of density pressure is also monotonically increasing. Therefore, this suggests that while matter particles accumulate during accretion around star models, the effective pressure contribution is likewise large enough to preserve stability and prevent collapse. 

\subsubsection*{Energy conditions}

\begin{figure}[bt]
\centering
\begin{minipage}{0.3\textwidth}
    \centering
    \includegraphics[width=\textwidth]{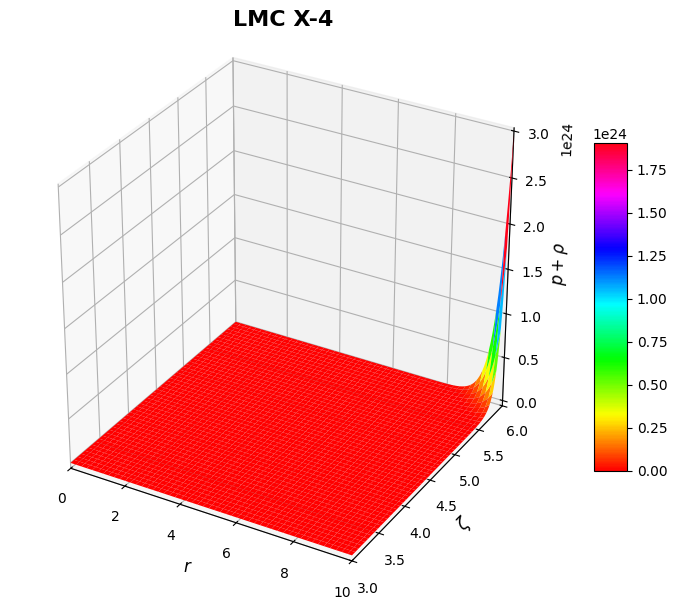}
    \\ \small (a)
\end{minipage}
\hfill
\begin{minipage}{0.3\textwidth}
    \centering
    \includegraphics[width=\textwidth]{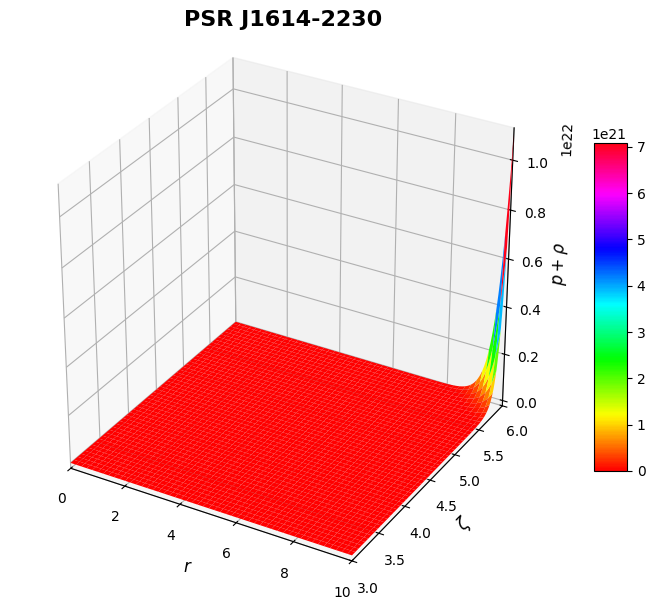}
    \\ \small (b)
\end{minipage}
\hfill
\begin{minipage}{0.3\textwidth}
    \centering
    \includegraphics[width=\textwidth]{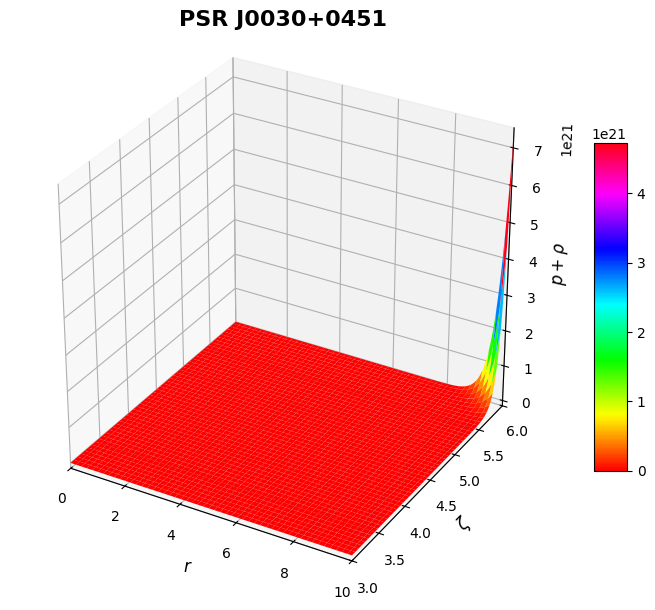}
    \\ \small (c)
\end{minipage}
\caption{ Evolution of the null energy conditions about radius $r$ for the power law model for the stellar models LMC X-4, PSR J1614-2230, and PSR J0030+0451. }
\label{f14}
\end{figure}

\begin{figure}[bt]
\centering
\begin{minipage}{0.3\textwidth}
    \centering
    \includegraphics[width=\textwidth]{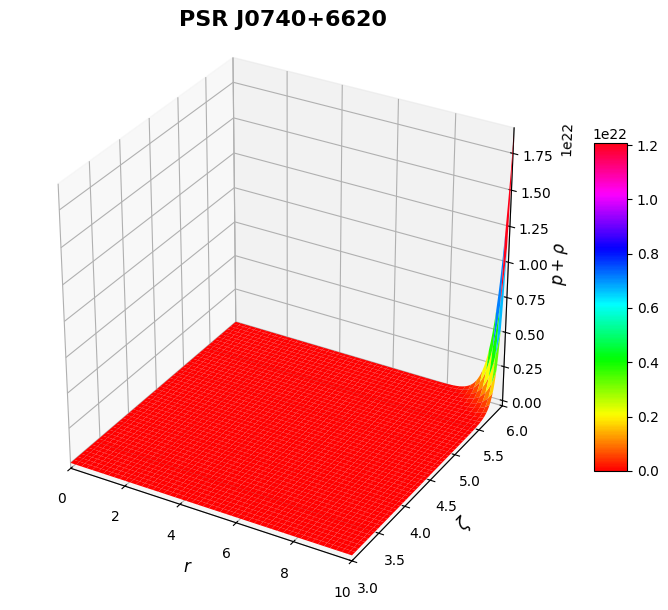}
    \\ \small (a)
\end{minipage}
\hfill
\begin{minipage}{0.3\textwidth}
    \centering
    \includegraphics[width=\textwidth]{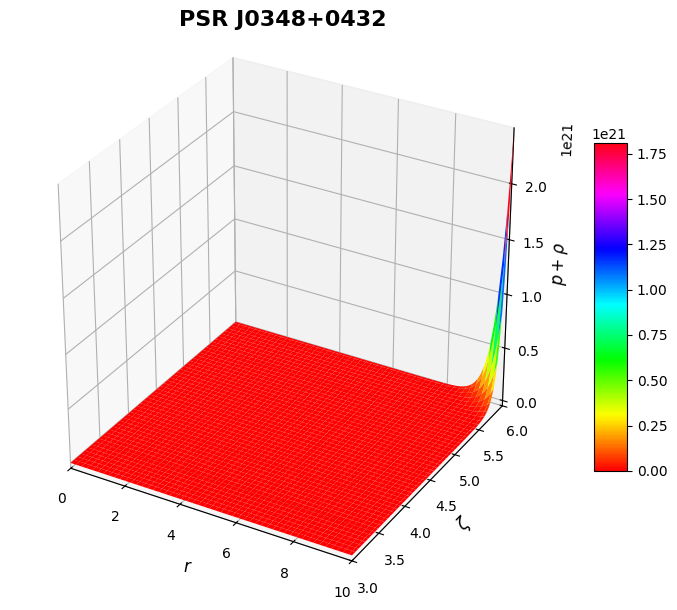}
    \\ \small (b)
\end{minipage}
\hfill
\begin{minipage}{0.3\textwidth}
    \centering
    \includegraphics[width=\textwidth]{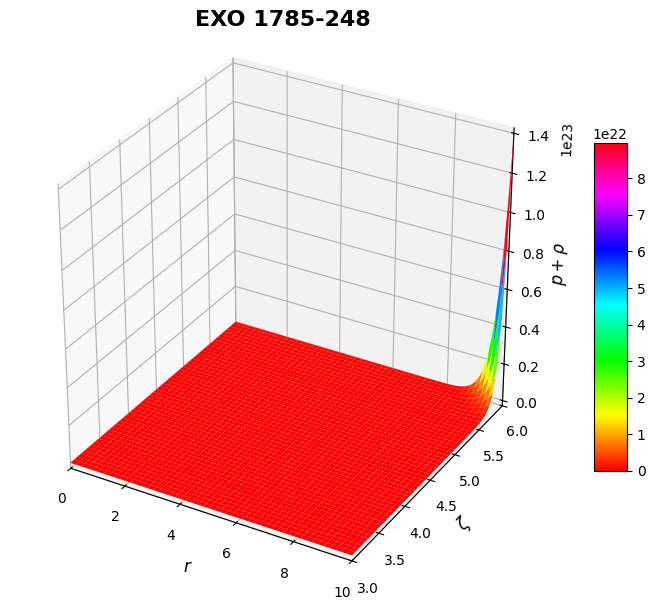}
    \\ \small (c)
\end{minipage}
\caption{ Evolution of the null energy conditions about radius $r$ for the power law model for the stellar models PSR J0348+0432, PSR J0348+0432 and EXO 1785-248. }
\label{f15}
\end{figure}

\begin{figure}[bt]
\centering
\begin{minipage}{0.3\textwidth}
    \centering
    \includegraphics[width=\textwidth]{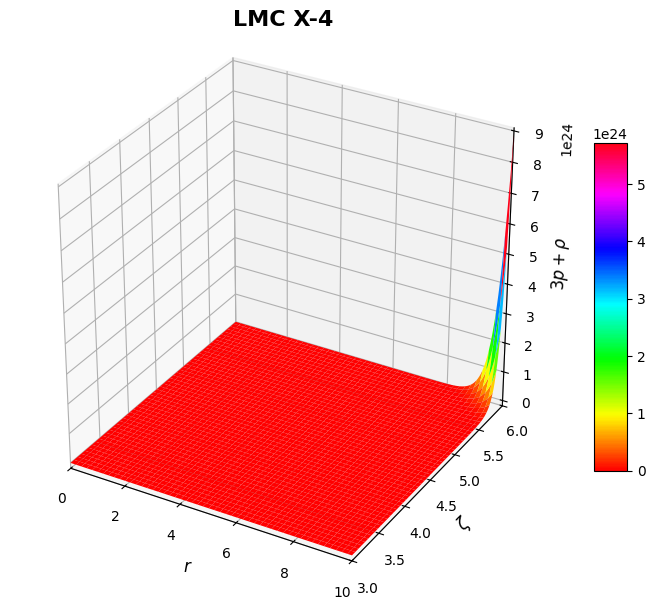}
    \\ \small (a)
\end{minipage}
\hfill
\begin{minipage}{0.3\textwidth}
    \centering
    \includegraphics[width=\textwidth]{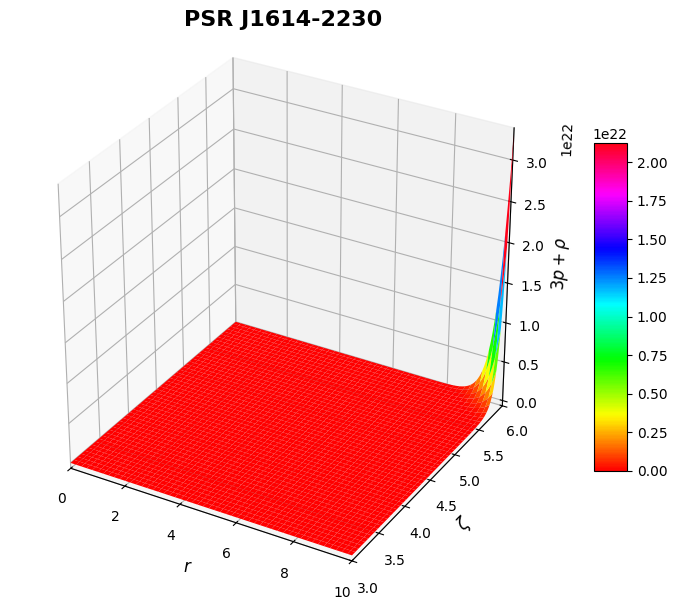}
    \\ \small (b)
\end{minipage}
\hfill
\begin{minipage}{0.3\textwidth}
    \centering
    \includegraphics[width=\textwidth]{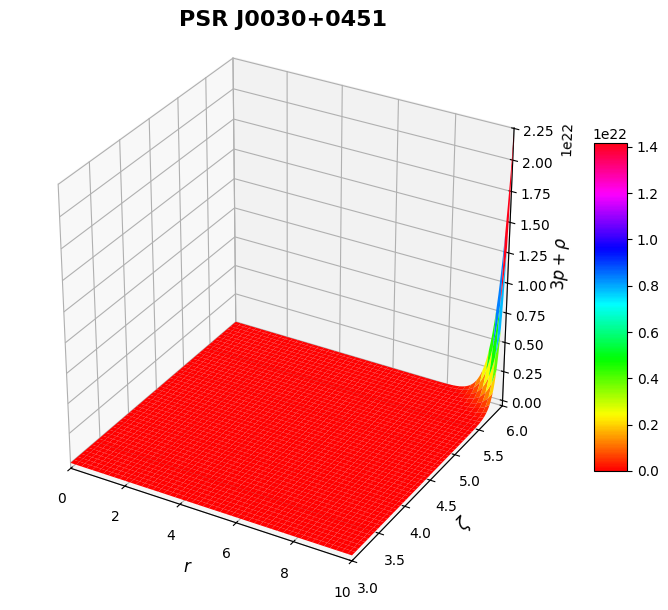}
    \\ \small (c)
\end{minipage}
\caption{ Evolution of the strong energy conditions about radius $r$ for the power law model for the stellar models LMC X-4, PSR J1614-2230, and PSR J0030+0451.  }
\label{f16}
\end{figure}

\begin{figure}[bt]
\centering
\begin{minipage}{0.3\textwidth}
    \centering
    \includegraphics[width=\textwidth]{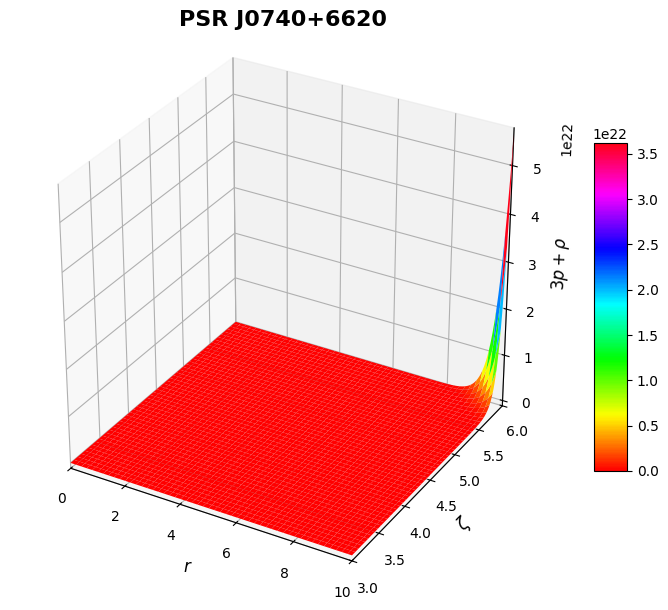}
    \\ \small (a)
\end{minipage}
\hfill
\begin{minipage}{0.3\textwidth}
    \centering
    \includegraphics[width=\textwidth]{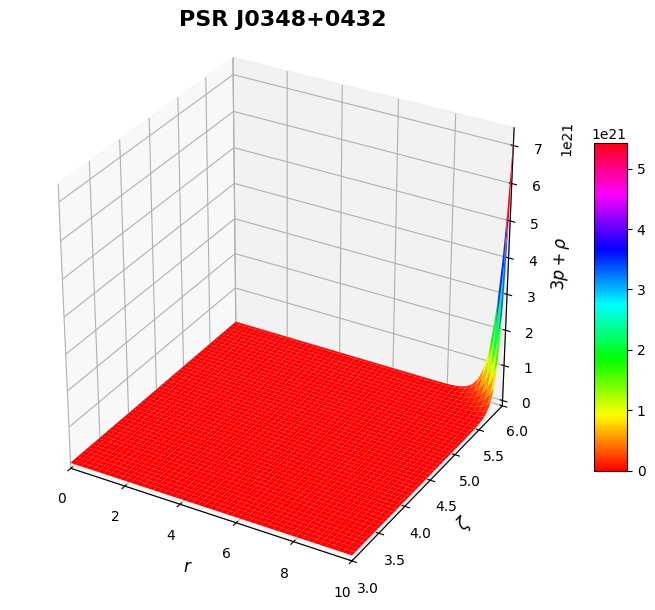}
    \\ \small (b)
\end{minipage}
\hfill
\begin{minipage}{0.3\textwidth}
    \centering
    \includegraphics[width=\textwidth]{sec5_power.png}
    \\ \small (c)
\end{minipage}
\caption{ Evolution of the strong energy conditions about radius $r$ for the power law model for the stellar models PSR J0348+0432, PSR J0348+0432 and EXO 1785-248. }
\label{f17}
\end{figure}

The evolution of the NEC is depicted in Fig. \ref{f14} and \ref{f15}, and SEC is depicted in Fig. \ref{f16} and \ref{f17}, which validates the energy conditions for our model. Here, we can observe that while the in Fig. \ref{f14} and \ref{f15} we demonstrate the NEC for the stellar models \emph{LMC X-4, PSR J1614-2230, PSR J0030+0451, PSR J0348+0432, PSR J0348+0432 and EXO 1785-248.}, the Fig. \ref{f16} and \ref{f17} confirms the SEC. Both the NEC and SEC show a monotonically growing pattern with respect to radius and exponent of the power law model (i.e., $\zeta$). The increasing NEC and SEC rates indicate that the gravitational source of the star increases as it gets closer to the core, creating greater internal stress to maintain stability and a steeper pressure gradient to resist gravity. Furthermore, the mass of the star distribution is more resilient to compression, leading to a stable compact stellar configuration concept.

\subsubsection*{Stability analysis}

\begin{figure}[bt]
\centering
\begin{minipage}{0.3\textwidth}
    \centering
    \includegraphics[width=\textwidth]{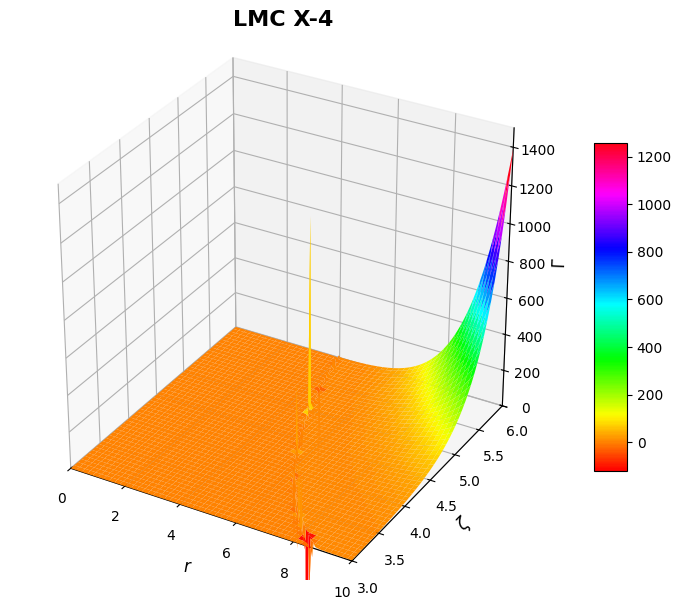}
    \\ \small (a)
\end{minipage}
\hfill
\begin{minipage}{0.3\textwidth}
    \centering
    \includegraphics[width=\textwidth]{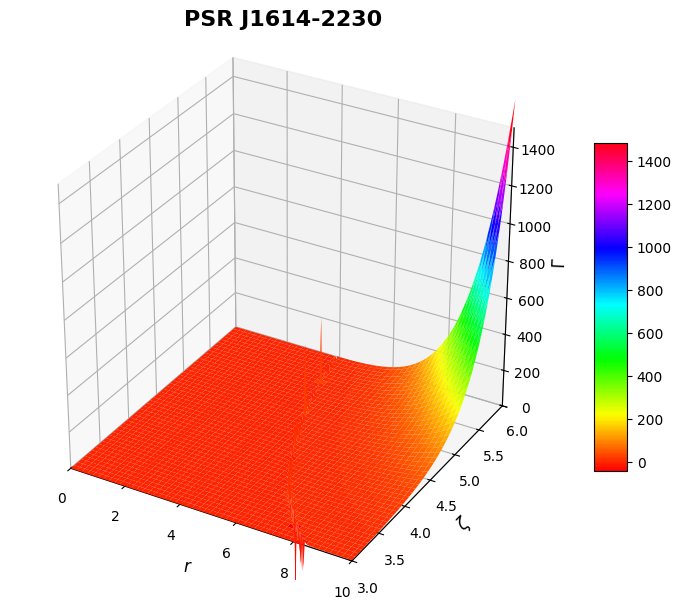}
    \\ \small (b)
\end{minipage}
\hfill
\begin{minipage}{0.3\textwidth}
    \centering
    \includegraphics[width=\textwidth]{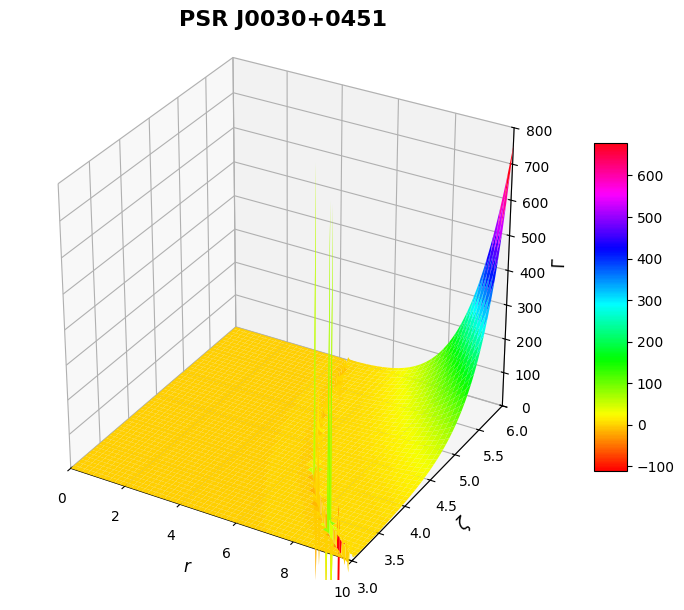}
    \\ \small (c)
\end{minipage}
\caption{ Evolution of Adiabatic Index about radius $r$ for the power law model for the stellar models LMC X-4, PSR J1614-2230, and PSR J0030+0451. }
\label{f18}
\end{figure}

\begin{figure}[bt]
\centering
\begin{minipage}{0.3\textwidth}
    \centering
    \includegraphics[width=\textwidth]{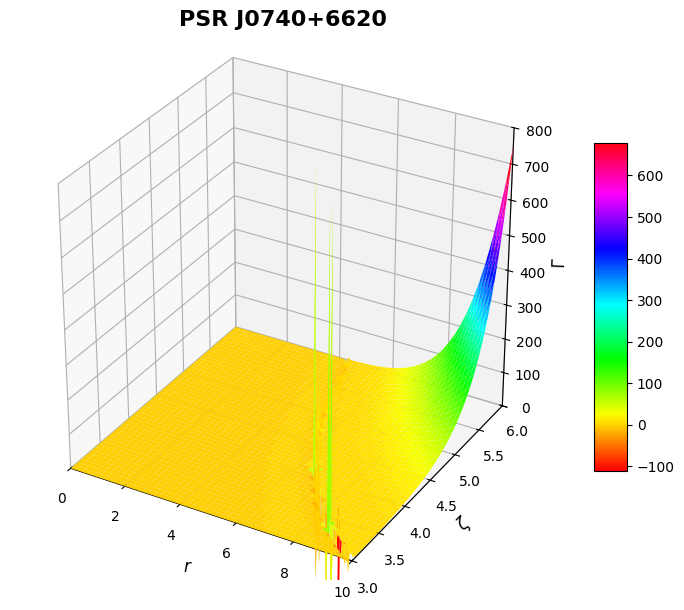}
    \\ \small (a)
\end{minipage}
\hfill
\begin{minipage}{0.3\textwidth}
    \centering
    \includegraphics[width=\textwidth]{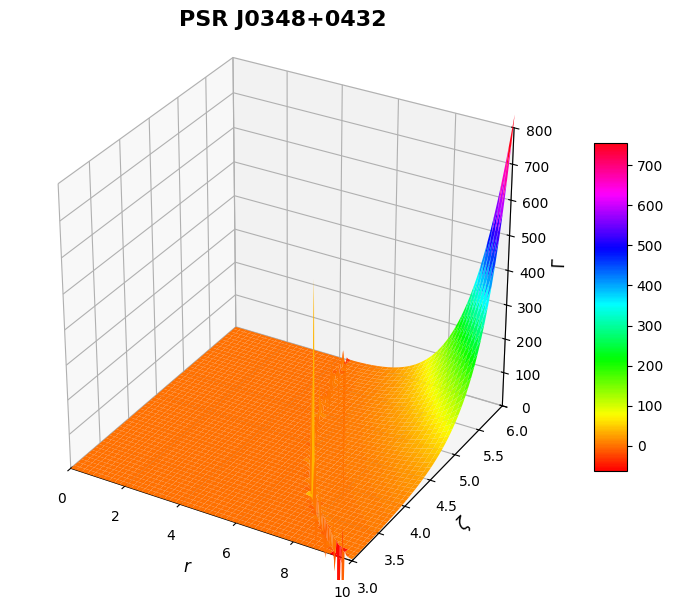}
    \\ \small (b)
\end{minipage}
\hfill
\begin{minipage}{0.3\textwidth}
    \centering
    \includegraphics[width=\textwidth]{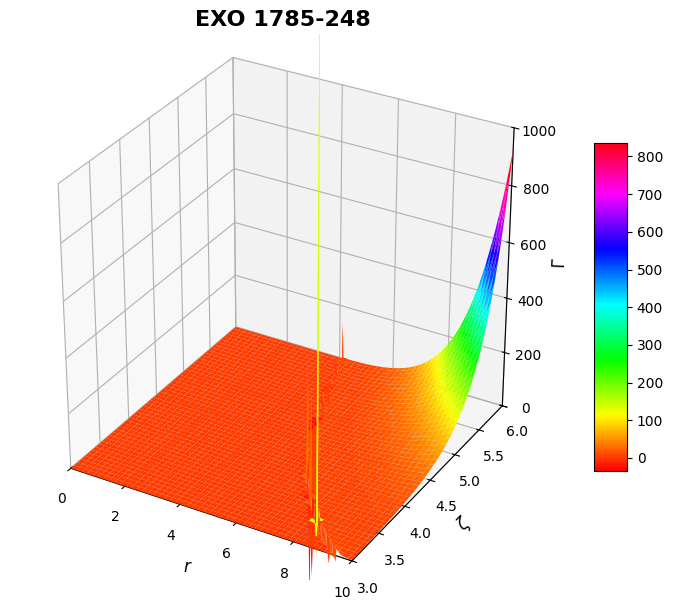}
    \\ \small (c)
\end{minipage}
\caption{ Evolution of Adiabatic Index about radius $r$ for the power law model for the stellar models PSR J0348+0432, PSR J0348+0432 and EXO 1785-248. }
\label{f19}
\end{figure}

In our study, Fig. \ref{f18} and \ref{f19} demonstrate the graphical representation of adiabatic index with respect to the radius and the exponent of the power law form (i.e., $\zeta$), for the star models \emph{LMC X-4, PSR J0740+6620, PSR J1614-2230, PSR J0348+0432, PSR J0030+0451, and EXO 1785-248}. The adiabatic index is crucial for understanding how changes in density affect the changes in pressure. Higher adiabatic index values show that pressure rises quickly with density, making the star model more resilient to compression and producing a more stable structure. The graph displays a monotonic increase in pattern within the interior of the stars, and the graphical depiction verifies that the adiabatic index for all compact stellar configurations lies within the permitted limit, which is $ > 4/3$. As density increases, the material enters a phase where nuclear forces strongly oppose further compression, supporting compact stellar configurations up to their maximum stable mass, as shown by an increase in $\Gamma$. An increasing adiabatic index means that pressure rises more sharply with density, the fluid or stellar matter is becoming stiffer, and the system is typically more resistant to gravitational collapse. Additionally, we can see that the adiabatic index increases more rapidly with higher values of exponent (i.e., $\zeta$).

\section{Concluding remarks}

In this study, we have evaluated the AdS star structure, which is a compact and equilibrium structure in the spacetime with a negative cosmological constant. At the conformal limit, the AdS star behaves like a radiation star. Also, noted that this is possible for a fermionic star or a Bose-condensate star at zero temperature. In particular, we have focused our study on the Boson star by considering the boson number density for $(d-2)$ dimensions in Eq. (\ref{E07}). Further noted that, we have established the relation between the Boson number density with the bulk density of state in Eq. (\ref{E10}), which guided us to evaluate the number of particles in Eq. (\ref{E11}) and the total mass in Eq. (\ref{E12}) of the AdS star, which is a special form of compact star. We have limited our study to the gravity in $AdS_5$, as authors in \cite{burikham2014comments} have shown that the other cases are qualitatively similar. Hence, for the $AdS_5$, the modified mass is evaluated for the star in Eq. (\ref{E16}). We have introduced a parameter $\alpha$ in Eq. (\ref{E15}), to establish the relation between chemical potential ($\mu$) and the bosonic mass ($m_B$) and the corresponding limitations on $\alpha$ have been established as well. In the next Section, we have considered bosonic mass ($m_B$) as an exponential function of radius, a quadratic function of radius, and a power law form of radius.  

We have considered the boson mass as an exponential function of radius in Eq. (\ref{E17}), and the consequent mass and compactness of the compact object are modified in Eq. (\ref{E18}) and Eq. (\ref{E32}), respectively. We have considered the bosonic mass as a quadratic function of radius in Eq. (\ref{E23}) in the next phase and the consequent mass and compactness of the compact object are evaluated further in Eq. (\ref{E24}) and Eq. (\ref{E34}), respectively. We have also chosen bosonic mass as a power law form in Eq. (\ref{E28}), and the corresponding estimation of the mass and compactness are demonstrated in Eq. (\ref{E29}) and Eq. (\ref{E30}).

The numerical computations for the mass, compactness and the surface redshift are described in Table \ref{T1} for exponential model background, Table \ref{T3} for quadratic model background, and Table \ref{T5} for power law model background, which confirms for choosing $\alpha$ as \emph{1.2, 1.45, 1.5, 1.38, 1.3, 1.23} we get a correspondence of masses for the compact objects \emph{LMC X-4, PSR J0740+6620, PSR J1614-2230, PSR J0348+0432, PSR J0030+0451, and EXO 1785-248} respectively. The compactness in the above-mentioned tables confirm that these are less than Buchdahl's limit, and the aforementioned stars are compact stellar configurations as well. In Fig. \ref{f1}, \ref{f5}, \ref{f5} and \ref{f9} we have graphically represented the mass-radius, compactness-radius, and surface redshift-radius for the exponential model, the quadratic model and the power law model, respectively. The mass-radius graphical representation in Fig. \ref{f1}, \ref{f5}, and \ref{f9} shows a monotonic increasing pattern with respect to radius, which is the accumulated mass distribution for the interior star structure for the exponential, quadratic and power law models, respectively. For the exponential, quadratic and power law models, the compactness is less Buchdahl's limit for the entire structure of the star and hence for the radius of about 13 km, the chances of the star collapsing and forming a black hole are negligible. Also, the numerical value of the compactness in Table \ref{T1}, \ref{T3}, and \ref{T5}, confirms that the considered stellar models are compact stellar configurations as the compactness lies within $0.1 \leq u \leq 0.25$ for the whole interior of the star.
The surface redshift has demonstrated an increasing pattern with respect to radius in Fig. \ref{f1}, \ref{f5} and \ref{f9}. The surface redshift explains the amount of light energy lost or redshifted during escape from the gravitational field. It is to be noted that increasing surface redshift indicates the gravitational field is stronger and the rate of photon particle emission from the outer surface is lower compared to the core. This can be concluded that the surface region possesses more gravitational potential compared to the core region, and hence the concentration of particle accumulation is more at the surface region, which indicates a denser shell for the stellar model and a less dense core. 

Now, using the coupled equation in Eq. (\ref{E04}), we have evaluated the density of the compact objects, and the pressure has been evaluated using standard thermodynamic relations in $(d-2)$ dimensions mentioned in Eq. (\ref{E21}). For three different assumptions of the bosonic mass functions, such as exponential form, quadratic form, and power law form, we have drawn three cases. The density, pressure and number of particles per unit volume are reconstructed for the exponential model in Eq. (\ref{E19}), Eq. (\ref{E20}), and Eq. (\ref{E22}), respectively, and the corresponding numerical values are described in Table \ref{T2}. The density, pressure and number of particles per unit volume are reconstructed for the quadratic model in Eq. (\ref{E25}), Eq. (\ref{E26}), and Eq. (\ref{E27}), respectively, and the corresponding numerical values are described in Table \ref{T4}. The density, pressure and number of particles per unit volume are reconstructed for the power law model in Eq. (\ref{E36}), Eq. (\ref{E37}), and Eq. (\ref{E38}), respectively, and the corresponding numerical values are described in Table \ref{T6}. It is to be noted that density reaches at its maximum for the exponential model for all the stars, and corresponding pressure and number of particles per unit volume are also maximum for the exponential model compared to the other two models. The graphical representation of density for the exponential model and the quadratic model is demonstrated in the left panel figures in Fig. \ref{f2} and \ref{f6}, respectively for the stellar models \emph{LMC X-4, PSR J0740+6620, PSR J1614-2230, PSR J0348+0432, PSR J0030+0451, and EXO 1785-248}. For the power-law model, the density is shown in Figs. \ref{f10} and \ref{f11} for the aforementioned stellar models. For all of the models, the density exhibits a monotonically increasing pattern. In contrast to the quadratic model, where the density continues to increase from the structure's core, the exponential model shows asymptotic flatness near the core, followed by a rigid ascending pattern. The thick shell of matter deposition around the surface in the stellar models is a result of accretion, as these graphics illustrate. In order to observe the sensitivity of the exponent for the density evolution, we have assessed the density evolution for the power law model with respect to radius and the exponent of the power law model. We can deduce that, though the density shows an increasing rate for this model as well but the increasing rate of density is more pronounced for higher exponent values. Therefore, the accretion surrounding the star structure is likewise supported by the power law model. The pressure is demonstrated in the right panel figures in Fig. \ref{f2} and \ref{f6} for the exponent model and the quadratic model, respectively. For the power-law model, the pressure is shown in Figs. \ref{f12} and \ref{f13}. For the exponential, quadratic, and power law models, the pressure evolution displays an erect pattern. In the quadratic model, the erection is monotonic after 3 km radius, while in the exponential model, a rigid erection occurs after 6 km radius. We have assessed the graphical demonstration in relation to the radius of the power law model and exponent, and the results show that pressure evolution is more evident for higher-order exponents. The accretion is supported by the increasing rate of pressure, just as the effective pressure contribution rises in conjunction with the increasing rate of density to preserve the equilibrium of the stellar structure. Therefore, the concentration of matter particles near the surface is unlikely to cause the structure to collapse. 

The AdS space has a time-like boundary at spatial infinity and functions as a gravity-confining box. Within the context of string theory, several authors \cite{gubser1996entropy,klebanov1997world,gubser1997absorption} have found evidence of the duality between gravitational physics in the bulk and gauge theory on the AdS space boundary. The space has a saddle-like form and negative curvature due to the negative cosmological constant. As the constraining gravity acts on spacetime like a potential wall, the AdS space draws everything back to the center instead of allowing the matter to escape to infinity. The gravity-confined structure traps matter and radiation in spacetime, which increases the chance of accretion around the star object. Contrary to flat spacetime, the accretion flow is stimulated in AdS space because the accretion energy cannot escape the boundary due to the geometry of the space, and the outer disk becomes hotter and denser, creating a more stable accretion flow. Our study supports this result of accretion for each case (i.e., the exponential model, the quadratic model, and the power law model), as the graphical demonstration illustrates. We will now examine the stability requirements for the previously specified cases in the sections. 

The requirements of energy conditions: $(i)$ $NEC : {\rho } +{p} \geq 0 $ , $(ii)$ $SEC : {\rho } + {p} \geq 0 $ , ${\rho } + 3 {p}  \geq 0$, must be satisfied in order for models of compact things to accurately represent the special behaviour of matter. Models must meet the energy limitations of general relativity in order to confirm the presence of non-exotic matter in star formations. The second rule of black hole thermodynamics and the Hawking-Penrose singularity theorems can be thoroughly examined in these conditions. For the energy-momentum tensor to consistently maintain positive throughout the stellar configuration, compact objects must satisfy the aforementioned inequalities. In Fig. \ref{f3}, and \ref{f7} the null energy condition and strong energy condition of aforementioned stellar models have been verified for exponential model and quadratic model respectively. The energy conditions for the power law model with respect to the radius and exponent of the power law model are supported by Figs. \ref{f14}, \ref{f15}, \ref{f16}, and \ref{f18}. We observe that the energy criteria are well satisfied and persists even for increasing values of the exponent. Hence there is no sign of existing non-exotic matter in the interior of the stellar models. The adiabatic index is crucial for determining if the formation of the pulsar is balanced in the presence of a tiny radial adiabatic influence. It should be noted that Bondi \cite{royal1869proceedings} claims that in this case, the model is unstable for $\Gamma $ less than $4/3$. Also, Chandrashekhar \cite{chandrasekhar1964dynamical} proposed this limiting value to evaluate the dynamical stability of anisotropic realistic stars with spherical symmetry in the presence of small perturbations in radial adiabatic. Heintzmann and Hillebrandt \cite{heintzmann1975neutron} state that for compact objects with a rising anisotropy factor, the relativistic stability requirement, $\Gamma $, should be greater than 4/3. Fig. \ref{f4} and \ref{f8} demonstrate the graphical evaluation of the adiabatic index for the exponential and the quadratic model, respectively. The steadily rising trend of the adiabatic index shown in Fig. \ref{f4} for the exponential model suggests that the rate at which pressure rises is more responsive to increases in density close to the surface. This implies that the stellar models exhibit greater resistance to gravitational collapse, resulting in the formation of more compact and stable objects.

We would like to note before winding up our work that we have theoretically investigated realistic compact stellar configurations in the AdS/CFT correspondence. Realistic equations of state that interpolate between low-density chiral-effective theory and high-density perturbative QCD are also produced by the ``top-down" models (V-QCD, D3-D7) and ``bottom-up" techniques, meeting the two solar-mass and tidal-deformability bounds. The Tolman-Oppenheimer-Volkoff (TOV) structure is reproduced by these holographic neutron-star solutions, which also predict that the deconfinement transition sets the maximum mass while the quark-matter phase is sufficiently soft to prevent stable quark cores \cite{hoyos2022holographic}. Previous research on degenerate fermionic stars in AdS provided the gravitational-collapse side of the story by demonstrating how a Chandrasekhar-type limit appears and how collapse to an AdS black hole correlates to thermalization in the dual gauge theory \cite{arsiwalla2011degenerate}. Hence, this argument supports that realistic compact stellar configuration can indeed be studied in AdS space with holographic dual. Since we have examined the condensation property of mass rather than the degeneracy property of pressure, we have substituted the bosonic gas for the ferimion gas in comparison to the previous study. A mass-radius comparison of compact stellar configuration was carried out by assuming that the bosonic mass is a function of radial coordinate in three distinct forms. Several models, including the power law, quadratic, and exponential models, have been studied in the context of physical characteristics including density, pressure, and mass. Compactness, surface redshift, adiabatic index, and energy conditions have all been studied for stability and equilibrium.

Therefore, we can deduce that in AdS gravity, compact objects would become accreted stars with denser outer disks. Additionally, the pressure will gradually grow to maintain equilibrium for the higher density outside the stellar structure. Although the generation of a black hole depends on the constraints of the radius value, this gravitational configuration creates a very compact object. The compactness of the stellar objects in the three scenarios shows that, although the structures are creating compact stellar configurations for the selected realistic stellar models, they may generate black holes for larger radii as the compactness increases with increasing radius. Also, the structures are well-stable and under cosmological perturbations.

\subsection{Data Availability}
The data sources have been duly acknowledged in the bibliography. 

\subsection{Funding Declaration}
The authors hereby declare that, the work reported here does not have any funding from any source.

\subsection*{Acknowledgements }
 Aroonkumar Beesham is grateful to the DSTI-NRF Centre of Excellence in Mathematical and
Statistical Sciences (CoE-MaSS), South Africa, for funding. Opinions expressed and conclusions arrived at are those of the authors and are not necessarily to be attributed to the CoE-MaSS.

\bibliography{sample}

\end{document}